%% file: main_plain.tex
\documentclass[11pt]{article}
\usepackage[letterpaper,margin=0.78in]{geometry}

\usepackage{amsmath,amssymb,mathtools}
\usepackage{xcolor}
\usepackage{tikz}
\usetikzlibrary{positioning,arrows.meta}
\usepackage{booktabs,tabularx,array}
\usepackage{cite}
\usepackage{microtype}
\usepackage{amsthm}
\usepackage[hidelinks]{hyperref}

\newcommand{\R}{\mathbb R}
\newcommand{\C}{\mathbb C}

\newcommand{\Ical}{\mathcal I}
\newcommand{\Jcal}{\mathcal J}
\newcommand{\Fcal}{\mathcal F}
\newcommand{\spec}{\operatorname{spec}}
\newcommand{\diag}{\operatorname{diag}}

\newtheorem{theorem}{Theorem}
\newtheorem{definition}{Definition}
\newtheorem{proposition}{Proposition}
\newtheorem{corollary}{Corollary}
\newtheorem{example}{Example}

\title{Information-Induced Conservatism Frontier of Distributed Stability Certificates}
\author{Xiaoyu Peng, Feng Liu
\footnote{The authors are with the Department of Electrical Engineering, Tsinghua University, Beijing, 100084, China.
(\texttt{pengxy19@tsinghua.org.cn}, \texttt{lfeng@mail.tsinghua.edu.cn})}
}
\date{}

\begin{document}
\maketitle
\input{text_info_struct}
\bibliographystyle{IEEEtran}
\bibliography{references}
\end{document}

%% file: text_info_struct.tex
\begin{abstract}
Conservatism in distributed stability certification is usually treated as a limitation of the certificates themselves. Therefore, to reduce the conservatism, researchers seek richer Lyapunov functions, tighter relaxations, or more powerful optimization methods. But this view misses a more fundamental limitation: a distributed certificate only sees the information exposed by its reporting interface. Once that information is compressed, physically different systems may become indistinguishable. If a stable system and a less stable or unstable system produce the same report, no sound method can separate them using that report. The resulting conservatism is caused by missing information, not by limitations of the analysis method. This paper asks where this fundamental information limit lies, and how much additional information is needed to move it. 

We model the available information as an information structure that partitions admissible systems into information fibers.
Within a fiber, every certificate is given identical information and therefore cannot distinguish between the different physical systems it contains.
Accordingly, the strongest sound certificate for a fiber is limited by the smallest stability margin across all systems in that fiber. This gives a method-independent benchmark. It also separates total conservatism into two parts: irreducible information-induced conservatism and additional method-induced conservatism. The framework then characterizes information order, task sufficiency, and the stability value of additional information. Three network--node studies expose the underlying mechanism. Network-spectrum compression loses the alignment between network modes and heterogeneous nodes. Scalar or eigenvalue-only passivity reports lose local directional alignment. Independent local certificates lose cross-node relations. These examples show that the critical missing information is often relational, not dimensional. 

We then move from diagnosis to information design. An information--conservatism frontier quantifies the smallest irreducible conservatism achievable under a prescribed budget. Its inverse problem gives the minimum information required for a desired conservatism. Through a task-relevant witness-pair characterization, we turn it into a weighted covering problem. 
A local sensitivity law gives a computable approximation and identifies which unresolved quantities are most valuable to reveal. The resulting framework tells whether conservatism should be reduced by improving the certificate or by enriching the information. 
\end{abstract}

    
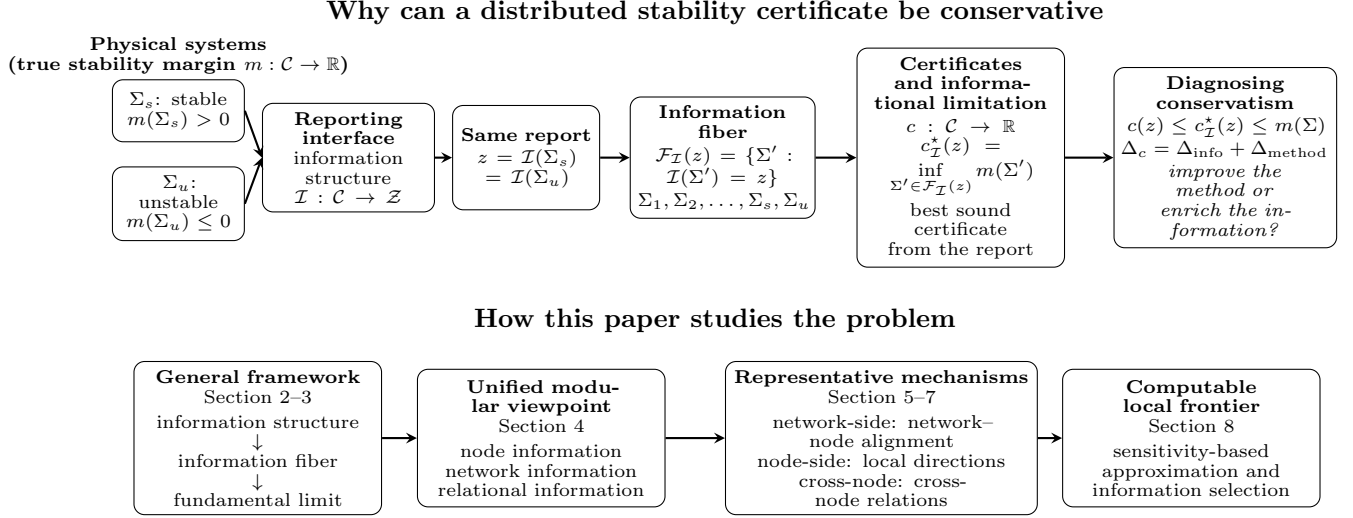
\begin{figure*}[htpb]
\centering
\begin{tikzpicture}[
    x=1cm,y=1cm,
    >=stealth,
    every node/.style={font=\fontsize{6.6}{7.4}\selectfont,text=black},
    box/.style={
        draw=black,
        rounded corners,
        align=center,
        inner sep=3.2pt,
        minimum height=1.30cm
    }
]


\node[font=\small\bfseries] at (7.95,4.25)
{Why can a distributed stability certificate be conservative};

\node[
    font=\fontsize{6.8}{7.4}\selectfont\bfseries,
    align=center
] at (0.85,3.68)
{Physical systems\\
(true stability margin $m:\mathcal C\rightarrow\mathbb R$)};

\node[box,text width=1.52cm,minimum height=0.82cm] (Ss) at (0.85,2.95)
{$\Sigma_s$: stable\\
$m(\Sigma_s)>0$};

\node[box,text width=1.52cm,minimum height=0.82cm] (Su) at (0.85,1.72)
{$\Sigma_u$: unstable\\
$m(\Sigma_u)\le0$};

\node[box,text width=2.05cm,minimum height=1.38cm] (I) at (3.10,2.33)
{\textbf{Reporting interface}\\
information\\
structure\\
$\Ical:\mathcal C\rightarrow\mathcal Z$};

\node[box,text width=1.72cm,minimum height=1.38cm] (z) at (5.45,2.33)
{\textbf{Same report}\\[0.4mm]
$z=\Ical(\Sigma_s)$\\
$=\Ical(\Sigma_u)$};

\node[box,text width=2.22cm,minimum height=1.38cm] (F) at (8.05,2.33)
{\textbf{Information fiber}\\[0.4mm]
$\Fcal_{\Ical}(z)=\{\Sigma':$\\
$\Ical(\Sigma')=z\}$\\[0.4mm]
$\Sigma_1,\Sigma_2,\ldots,\Sigma_s,\Sigma_u$};

\node[box,text width=2.52cm,minimum height=1.38cm] (Cstar) at (11.20,2.33)
{\textbf{Certificates and informational limitation}\\[0.4mm]
$c: \mathcal C\rightarrow\mathbb R$
\\
$c_{\Ical}^{\star}(z)
=\displaystyle\inf_{\Sigma'\in\Fcal_{\Ical}(z)}m(\Sigma')$\\[0.4mm]
best sound certificate\\
from the report};

\node[box,text width=2.72cm,minimum height=1.38cm] (Diag) at (14.70,2.33)
{\textbf{Diagnosing conservatism}\\[0.4mm]
$c(z)\le c_{\Ical}^{\star}(z)\le m(\Sigma)$\\[0.4mm]
$\Delta_c=\Delta_{\rm info}+\Delta_{\rm method}$\\[0.4mm]
\textit{improve the method or}\\
\textit{enrich the information?}};

\draw[->,thick] (Ss.east) -- (I.west);
\draw[->,thick] (Su.east) -- (I.west);
\draw[->,thick] (I.east) -- (z.west);
\draw[->,thick] (z.east) -- (F.west);
\draw[->,thick] (F.east) -- (Cstar.west);
\draw[->,thick] (Cstar.east) -- (Diag.west);


\node[font=\small\bfseries] at (7.95,0.18)
{How this paper studies the problem};

\node[box,text width=3.05cm,minimum height=1.82cm] (B1) at (1.90,-1.38)
{\textbf{General framework}\\
Section~\ref{sec:information-limit}--\ref{sec:equivalence-limits}\\[0.7mm]
information structure\\
$\downarrow$\\
information fiber\\
$\downarrow$\\
fundamental limit};

\node[box,text width=3.05cm,minimum height=1.82cm] (B2) at (5.65,-1.38)
{\textbf{Unified modular viewpoint}\\
Section~\ref{sec:distributed-structure}\\[0.7mm]
node information\\
network information\\
relational information};

\node[box,text width=3.90cm,minimum height=1.82cm] (B3) at (10.15,-1.38)
{\textbf{Representative mechanisms}\\
Section~\ref{sec:network-spectrum-insufficiency}--%
\ref{sec:cross-node-info-value}\\[0.7mm]
network-side: network--node alignment\\
node-side: local directions\\
cross-node: cross-node relations};

\node[box,text width=3.18cm,minimum height=1.82cm] (B4) at (14.25,-1.38)
{\textbf{Computable local frontier}\\
Section~\ref{sec:quantification}\\[0.7mm]
sensitivity-based\\
approximation and\\
information selection};

\draw[->,thick] (B1.east) -- (B2.west);
\draw[->,thick] (B2.east) -- (B3.west);
\draw[->,thick] (B3.east) -- (B4.west);

\end{tikzpicture}

\caption{
The lower panel demonstrate the general idea.
A reporting interface induces an information structure under which physically different systems may produce the same report and therefore belong to the same information fiber.
The best sound certificate attainable from that report is the worst-case stability margin over the fiber, providing the information-limited benchmark $c_{\Ical}^{\star}$.
Comparison with a particular certificate separates total conservatism into information-induced and method-induced components and indicates whether further improvement should target the certificate method or the reporting interface.
Throughout the paper, \textit{information} is used in this task-specific distinguishability sense, rather than in the Shannon sense. A higher-dimensional report can have little or even zero stability value if it does not separate the systems that determine the worst case.
\\
The lower panel summarizes the progression from the general framework to the modular interpretation, representative mechanisms of information-induced conservatism, and the computable local information frontier. Besides, it shows their positions in the paper.
}
\label{fig:paper-overview}
\end{figure*}

\section{Introduction}\label{sec:introduction}

Distributed and compositional stability certification is a recurring challenge in large-scale interconnected systems, where assembling and repeatedly analyzing a complete closed-loop model can be computationally expensive or impractical.
Rather than reconstructing the entire model, distributed approaches seek local conditions whose satisfaction guarantees stability. Power-system stability studies provide prominent examples, including small-phase/passivity-based \cite{Yang_Distributed_2020, He_Passivity_2024, Haberle_Decentralized_2026}, small-gain-based \cite{Lin_Small_2026}, gain--phase mixed \cite{Huang_GainPhase_2024}, and integral-quadratic-constraint-based methods \cite{Liu_Scalable_2026}.
The broader architectural idea is not specific to power systems: the certificate operates on selected component and interconnection descriptions rather than on the complete physical system.
This compression makes scalable analysis possible, but it can also make the resulting certificates conservative.
\emph{There are two fundamentally different sources of conservatism. The analysis method may fail to exploit the available information, or the information itself may be insufficient for the stability task.}

Much of the recent progress has focused on making the resulting certificates less conservative. Passivity restrictions have been relaxed through lifting the scalar passivity indices to matrix-valued descriptions to retain directional and inter-channel information \cite{Ru_MatrixPassivity_2026,Ru_Less_2026}. Multiplier selection for broad classes of decentralized frequency-domain certificates has been formulated as a convex semidefinite program, enabling systematic searches for the least-conservative certificate within a prescribed family \cite{Gorbunov_SDP_2026}. These developments address an important methodological question: \emph{how effectively does a certificate use the information supplied to it?} They do not, however, answer the logically prior question: \emph{how much can any sound certificate accomplish with that information in the first place?} In the extreme case, if a stable and an unstable system are indistinguishable under the available information, no improvement of the certificate algorithm alone can certify the stable one.

 This distinction is operational rather than merely conceptual. When a distributed certificate is conservative, one may invest in a richer Lyapunov construction, a larger multiplier family, a tighter relaxation, or a more expensive optimization. Such effort is justified only if the existing information still permits a better sound certificate. If the best certificate compatible with the current reporting interface is already restrictive, algorithmic refinement cannot recover the missing margin.
 Instead, the reporting interface itself must be enriched. The central purpose of this paper is therefore to provide a benchmark that tells these two situations apart.

A simple example makes the issue concrete. 
Suppose a distributed certificate knows the individual node dynamics and the network Laplacian spectrum.
Imagine a star-shaped interconnection: one hub is connected to all the other nodes. Suppose all subsystems have the same dynamics except one with a stronger dynamic gain. Compare two cases: this high-gain subsystem is at the hub in the first case and at a leaf in the second. Both networks are the same star and therefore have exactly the same network spectrum. But the high-gain subsystem interacts with the rest of the system very differently at the hub than at a leaf. For a suitable coupling strength and a class of node dynamics, one placement can remain asymptotically stable while the other fails the Hurwitz stability test. A certificate that knows the individual node models but receives only the network eigenvalues sees the same report in both cases, so it cannot soundly distinguish the stable system from its {unstable counterpart}. The missing information is not another eigenvalue; it is how the node dynamics are placed relative to the network's critical directions. This example illustrates a possible information insufficiency and will be further clarified in Section~\ref{sec:network-spectrum-insufficiency}.

 This distinction turns conservatism into an information-sufficiency problem. Blackwell's comparison of experiments provides a classical way of ordering information structures according to whether one can be obtained from another by garbling \cite{Blackwell_Equivalent_1953}. Alternatively, the data-informativity framework asks whether available data are sufficient for a specified analysis or control task over all systems consistent with those data \cite{vanWaarde_Data_2020,vanWaarde_Informativity_2023}. In distributed stability analysis, however, the relevant object is not necessarily a data set, but the information structure induced by the reporting interface itself. A network spectrum, a scalar or matrix-valued component index, or a collection of independent local descriptions may identify physically different interconnected systems as identical from the certificate's viewpoint. \emph{The fundamental benchmark is therefore not another sufficient stability condition, but the best stability statement that can be made by any sound method restricted to the same information.} To the best of our knowledge, this information-limited optimum and the resulting separation between information-induced and method-induced conservatism have not been characterized for distributed stability certificates in the literature.

 A related algorithm-independent viewpoint appears in information-based complexity, where the radius of information quantifies the irreducible worst-case error associated with restricted information structures \cite{Traub_Perspectives_1992}. The present setting differs from the standard information-based-complexity setting in two respects: soundness imposes a one-sided lower-certificate requirement, and the admissible information structures are dictated by modular network--node reporting interfaces rather than by arbitrary measurements. This motivates optimizing not over unrestricted structures, but over physically meaningful reporting-interface settings.


This paper develops an information-based framework for answering this question. The framework is stated for general classes of interconnected systems and task-dependent stability margins.
Its contributions follow a benchmark--diagnosis--design logic.

\begin{enumerate}
\item \textbf{Benchmark: the best certificate permitted by fixed information.}
We define the pointwise largest sound certificate as the worst-case margin over all systems compatible with the reported information. This gives a method-independent limit before any particular Lyapunov function, multiplier, or numerical algorithm is chosen.
\item \textbf{Diagnosis: separating information- and method-induced conservatism.} The benchmark decomposes total conservatism into information-induced conservatism and method-induced conservatism. Information sufficiency is characterized on the induced information fibers, with stable/unstable pairs in the same fiber serving as constructive impossibility witnesses. 
\item \textbf{Design: identifying stability-relevant information.} We order information structures and define a budgeted information--conservatism frontier over admissible refinements. Network-, node-, and cross-node examples expose the lost relations, while witness pairs, an exact passivity-direction frontier, and local sensitivity provide mechanisms for selecting useful additional information.
\end{enumerate}
Together, these results turn conservatism analysis into a decision problem: \emph{whether the next improvement should target the certificate method or the information structure}.
Fig.~\ref{fig:paper-overview} summarizes the central question and the organization of the paper.


 The remainder of the paper is organized as follows. Section~\ref{sec:information-limit} introduces information structures and the optimal certificate under fixed information. Section~\ref{sec:equivalence-limits} develops the corresponding equivalence-class, information-sufficiency, and fundamental-limit results. It further defines the budgeted information--conservatism frontier over admissible reporting-interface refinements and gives a witness-pair characterization of minimum sufficient information. Section~\ref{sec:distributed-structure} explains the network--node decomposition and unifies the sources of information-induced conservatism. Sections~\ref{sec:network-spectrum-insufficiency}--\ref{sec:cross-node-info-value} then study network-side, node-side, and cross-node information-induced conservatism, respectively. {Section~\ref{sec:quantification} develops a computable local approximation of the information frontier from stability sensitivity, including a budgeted first-order information-selection law, and Section~\ref{sec:conclusion} concludes the paper.}

\textbf{Notation}: The real and complex number sets are denoted by $\R$ and $\C$.
For a matrix $M$, $M^H$ denotes its Hermitian transpose, $\lambda(M)$ its eigenvalue, $\spec(M)$ its spectrum, and
$\alpha(M)=\max\Re(\spec(M))$ the largest real part of its eigenvalues.
The set of $n$-dimensional symmetric matrices is represented by $\mathbb S^{n}$.

\section{Information Structure and Value for Stability Certification}
\label{sec:information-limit}

 This section builds the conceptual framework required by the certificate question. We first specify the \emph{ground-truth task margin} $m$, and then specify \emph{what the distributed certificate can observe} through the information structure $\Ical$. A concrete method is represented by a certificate map $c$ acting only on that report. Soundness compares $c$ with the ground truth and leads to the information-limited optimum $c_{\Ical}^{\star}$, which is the best certificate attainable without changing the information structure. The gap between $m$ and $c_{\Ical}^{\star}$ is then the irreducible information-induced conservatism $\Delta_{\rm info}$. Finally, ordering and refinement of information structures quantify how this limit changes when additional information is supplied. This sequence is the organizing line used throughout the remainder of the paper: \emph{task $\rightarrow$ information $\rightarrow$ sound certificate $\rightarrow$ information-limited optimum $\rightarrow$ conservatism}

\subsection{Stability Evaluation Task and Information Structure}

Consider a class of admissible interconnected systems $\Sigma\in\mathcal C$, where
$\Sigma$ contains the complete information about the node dynamics, network
topology, parameters, and operating point relevant to the stability task.

\begin{definition}[Ground-Truth Stability Margin]
\label{def:ground-truth-margin}
A specified stability-evaluation task is represented by a real-valued map
\begin{align}
\boxed{m:\mathcal C\rightarrow\R,\qquad \Sigma\mapsto m(\Sigma)},
\label{eq:margin-general}
\end{align}
where $m(\Sigma)>0$ if and only if $\Sigma$ is stable.
The quantity $m(\Sigma)$ is called the \emph{ground-truth stability margin} of
the complete system $\Sigma$.
\end{definition}

The particular choice of $m$ is task-dependent. For small-signal stability, a
natural choice is the negative closed-loop spectral abscissa. Other margins,
such as the distance to a prescribed $\mathcal D$-stability boundary or an
admissible coupling margin, can be used without changing the framework below.

 \begin{definition}[Information Structure and Information Fiber] 
\label{def:information-structure}
\label{def:information-fiber}
A distributed or compositional stability analysis is assumed to observe the
 complete system only through an information structure 
\begin{equation}
\boxed{\Ical:\mathcal C\rightarrow\mathcal Z,\qquad z=\Ical(\Sigma)},
\label{eq:information-map}
\end{equation}
where $\Ical$ is called the \emph{information structure}. For a reported observation
 $z\in\mathcal Z$, define the corresponding \emph{information fiber} as 
\begin{equation}
\Fcal_{\Ical}(z)
:=
\left\{\Sigma'\in\mathcal C:\Ical(\Sigma')=z\right\}.
\label{eq:information-fiber}
\end{equation}
 When the report is generated by a particular system $\Sigma$, we use the shorthand 
 \begin{equation} 
 \Fcal_{\Ical}(\Sigma) 
 := 
 \Fcal_{\Ical}(\Ical(\Sigma)). 
 \label{eq:information-fiber-system-shorthand} 
 \end{equation} 
\end{definition}

 The information structure $\Ical$ may collect local node reports, network or operating-state 
 reports, selected model coefficients, or any other quantities available to  
the certificate. All systems in $\Fcal_{\Ical}(z)$ are indistinguishable to any
method whose input is restricted to $\Ical$. Thus, $\Ical$ partitions the
  physical system class into information fibers, and stability certification  
  is fundamentally a task over these fibers rather 
than over individual models. This view is closely related to data
informativity \cite{vanWaarde_Data_2020}. The induced equivalence-class
structure and its implications for fundamental conservatism are developed
formally in Section~\ref{sec:equivalence-limits}.

\begin{example}[Passivity-Based Stability Evaluation]
\label{ex:passivity-information}
Passivity provides a concrete example for the proposed methodology, which will be reused for illustration throughout the paper. In a
standard passivity-based distributed assessment, each node is represented by
a local passivity index, while the network is represented by a complementary
description. A compositional interconnection condition then combines these
local reports to certify the closed-loop system without requiring the certificate
to reveal detailed node models. Let the matrix passivity index
\cite{Ru_Less_2026} of the $i$-th node be
$\Pi_i\in\mathbb S^2$ and the network index be $\Pi_g\in\mathbb S^{2n}$.
Define $\Pi=\operatorname{blkdiag}(\Pi_1,\ldots,\Pi_n)$; a typical condition
then takes the form $\Pi-\Pi_g\succ0$. Hence the passivity method operates on
compressed information $\Pi$ rather than on the complete system $\Sigma$.

 A legacy scalar reporting interface further compresses each matrix report to 
$\pi_i=\lambda_{\min}(\Pi_i)$ \cite{Peng_Compositional_2026}. The scalar report
merges all nodes sharing the same weakest-direction index. Matrix-valued
passivity retains precisely part of the information discarded by this
 scalarization. Denote the scalar and matrix passivity information structures by $\Ical_{\rm sc}^{\Pi}$ and $\Ical_{\rm mat}^{\Pi}$, respectively. 
\end{example}

\subsection{Stability Certificates, Soundness, and Optimality}

\begin{definition}[Stability-Margin Certificate]
\label{def:certificate-map}
For a fixed information structure $\Ical$, a particular stability-analysis
method using only the reported information $z=\Ical(\Sigma)$ is represented by
a map
\begin{equation}
\boxed{c:\mathcal Z\rightarrow\R,\qquad z\mapsto c(z)},
\label{eq:certificate-map}
\end{equation}
where $c(z)$ is the stability margin certified by that method, expressed in
the same margin convention as $m$. A positive value $c(z)>0$ certifies
stability, whereas $c(z)\le0$ means only that the method cannot certify a
positive margin from the available information.
\end{definition}

\begin{example}
The certificate $c$ may be obtained from a spectral criterion, a passivity
criterion, or any other stability-analysis method, provided that its output is
expressed in the same stability-margin convention as $m$.
\end{example}

The three basic objects introduced above have distinct roles:
$m$ evaluates the true margin from the complete system, $\Ical$ specifies what
information is available to a distributed certificate, and $c$ maps that
available information to a certified margin. Equivalently, the complete and
distributed evaluation paths are
\begin{equation}
\Sigma\xrightarrow{\ m\ }m(\Sigma),
\qquad
\Sigma\xrightarrow{\ \Ical\ }z
\xrightarrow{\ c\ }c(z).
\label{eq:two-evaluation-paths}
\end{equation}

\begin{definition}[Sound Stability-Margin Certificate]
\label{def:sound-certificate}
A certificate map $c:\mathcal Z\rightarrow\R$ is \emph{sound} if
\begin{equation}
c(\Ical(\Sigma))\le m(\Sigma),
\qquad \forall\Sigma\in\mathcal C.
\label{eq:sound-certificate}
\end{equation}
A sound certificate may underestimate the true margin, but it cannot
overestimate it.
\end{definition}

\begin{theorem}[Optimal Certificate under Fixed Information Structure]
\label{thm:optimal-certificate}
 For a fixed information structure $\Ical$, the pointwise largest sound certificate is 
\begin{equation}
\boxed{
c_{\Ical}^{\star}(z)
=
\inf_{\Sigma\in\Fcal_{\Ical}(z)}
m(\Sigma)}.
\label{eq:optimal-certificate}
\end{equation}
\end{theorem}

\begin{proof}
Let $c$ be any sound certificate. For every
$\Sigma\in\Fcal_{\Ical}(z)$, \eqref{eq:sound-certificate} gives
$c(z)\le m(\Sigma)$, and hence
\begin{equation}
c(z)\le\inf_{\Sigma\in\Fcal_{\Ical}(z)}m(\Sigma).
\end{equation}
Conversely, the right-hand side of \eqref{eq:optimal-certificate} does not exceed $m(\Sigma)$ for any system in the same information fiber and is 
therefore sound. Thus it is the pointwise optimum among all sound certificates.
\end{proof}

When evaluated at a particular system, we use the shorthand
$c_{\Ical}^{\star}(\Sigma):=
c_{\Ical}^{\star}(\Ical(\Sigma))$. Theorem~\ref{thm:optimal-certificate}
immediately gives, for every sound certificate $c$,
\begin{equation}
c(\Ical(\Sigma))
\le
c_{\Ical}^{\star}(\Ical(\Sigma))
\le
m(\Sigma).
\label{eq:certificate-hierarchy}
\end{equation}
Equation~\eqref{eq:certificate-hierarchy} is the central hierarchy of the
framework: $m$ is the oracle margin, $c_{\Ical}^{\star}$ is the best margin
permitted by the available information, and $c$ is the margin delivered by a
particular sound method.

The significance of Theorem~\ref{thm:optimal-certificate} is independent of
the chosen analysis technique, such as a particular Lyapunov function,
passivity condition, LMI, or numerical algorithm. \emph{Even an arbitrarily powerful
certificate cannot exceed $c_{\Ical}^{\star}$ without using information that
 further refines the current information fiber.} 

For a particular sound method $c$, its pointwise conservatism is
\begin{equation}
\Delta_c(\Sigma;\Ical)
:=
m(\Sigma)-c(\Ical(\Sigma))
\ge0.
\label{eq:method-gap}
\end{equation}

 \begin{definition}[Information-Induced Conservatism] 
\label{def:information-gap}
For the true system $\Sigma$, the irreducible conservatism caused solely by the
information structure $\Ical$ is
\begin{equation}
\boxed{
\Delta_{\rm info}(\Sigma;\Ical)
:=
m(\Sigma)-c_{\Ical}^{\star}(\Ical(\Sigma))}.
\label{eq:information-gap}
\end{equation}
\end{definition}

 For a particular sound method $c$, define its \emph{method-induced conservatism} as 
 \begin{equation} 
 \Delta_{\rm method}(\Sigma;\Ical) 
 := 
 c_{\Ical}^{\star}(\Ical(\Sigma))-c(\Ical(\Sigma)) 
 \ge0. 
 \label{eq:method-induced-conservatism} 
 \end{equation} 

 Combining \eqref{eq:method-gap}--\eqref{eq:method-induced-conservatism} gives 
\begin{equation}
\boxed{\begin{aligned}
\Delta_c(\Sigma;\Ical)
 ={}&\underbrace{m(\Sigma)-c_{\Ical}^{\star}(\Ical(\Sigma))}_{\Delta_{\rm info}(\Sigma;\Ical)}+\underbrace{c_{\Ical}^{\star}(\Ical(\Sigma))-c(\Ical(\Sigma))}_{\Delta_{\rm method}(\Sigma;\Ical)} 
\end{aligned}}
\label{eq:conservatism-decomposition}
\end{equation}
where both terms on the right-hand side are nonnegative. \emph{Thus
 $\Delta_{\rm info}$ is the information-induced conservatism that no sound method can 
 remove without additional information, whereas $\Delta_{\rm method}$ is the 
 method-induced conservatism measuring how far a particular method is from the 
 information-limited optimum.}

The hierarchy can be read as the short chain
\[
\Sigma
\ \xrightarrow{\Ical}\
z
\ \xrightarrow[\text{best possible}]{c_{\Ical}^{\star}}\
c_{\Ical}^{\star}(z)
\ \le\
m(\Sigma),
\qquad
c(z)\le c_{\Ical}^{\star}(z),
\]
which separates two questions that are otherwise easy to conflate: whether the \emph{information} is sufficient, and whether the \emph{chosen method} uses that information well. The upper panel of Fig.~\ref{fig:paper-overview} visualizes this hierarchy
together with its decisive failure mode: two systems on different sides of the stability boundary that collapse to the same report.

\subsection{Information Order and Stability Value}

\begin{definition}[Deterministic Informativeness Order]
\label{def:information-order}
If there exists a deterministic map $f$ such that
\begin{equation}
\Ical_1=f\circ\Ical_2,
\label{eq:garbling}
\end{equation}
we write $\Ical_2\succeq\Ical_1$ and say that $\Ical_2$ is at least as
informative as $\Ical_1$ for distributed stability evaluations.
\end{definition}

This is the deterministic counterpart of the Blackwell garbling order
\cite{Blackwell_Equivalent_1953}. Strictly, $\succeq$ is a preorder on
 information structures. After quotienting out information structures that can be deterministically 
 recovered from each other, it becomes a partial order on the resulting 
 information structures. Moreover, if $\Ical_2\succeq\Ical_1$, every 
$\Ical_2$-fiber is contained in the corresponding $\Ical_1$-fiber. Therefore,
for any $\Sigma\in\mathcal C$,
\begin{equation}\label{eq:cert-order}
    c_{\Ical_2}^{\star}(\Ical_2(\Sigma))
\ge
c_{\Ical_1}^{\star}(\Ical_1(\Sigma)),\;
\Delta_{\rm info}(\Sigma;\Ical_2)
\le
\Delta_{\rm info}(\Sigma;\Ical_1).
\end{equation}

\begin{example}[Scalar-to-Matrix Passivity Refinement]
\label{ex:passivity-refinement}
 Returning to Example~\ref{ex:passivity-information}, the scalar report is a 
 deterministic function of the matrix report through the minimum-eigenvalue 
map. Therefore
\begin{equation}
 \Ical_{\rm mat}^{\Pi}\succeq\Ical_{\rm sc}^{\Pi}, 
\qquad
 \Fcal_{\Ical_{\rm mat}^{\Pi}}(\Sigma) 
\subseteq
 \Fcal_{\Ical_{\rm sc}^{\Pi}}(\Sigma). 
\label{eq:matrix-scalar-fiber-order}
\end{equation}
For small-signal stability, choose the task margin
$m(\Sigma)=m_\alpha(\Sigma):=-\alpha(\Sigma)$, where
$\alpha(\Sigma)$ is the closed-loop spectral abscissa. Then
\begin{equation}
 c_{\Ical_{\rm mat}^{\Pi}}^{\star}(\Ical_{\rm mat}^{\Pi}(\Sigma)) 
\ge
 c_{\Ical_{\rm sc}^{\Pi}}^{\star}(\Ical_{\rm sc}^{\Pi}(\Sigma)). 
\label{eq:passivity-certificate-order}
\end{equation}
Thus, a reduction in conservatism from scalar to matrix passivity information
can be interpreted without first selecting a particular matrix LMI: the matrix
report removes some adverse directional systems that remain compatible with
the scalar report. Whether a specific certificate $c$ fully exploits this
additional information is a separate question; Section~\ref{sec:passivity-information-value}
 quantifies this information-induced conservatism explicitly. 
\end{example}

\begin{example}[$x$--$z$ Graph and SRG: Different Representations of the Same Information]
\label{ex:xz-srg-equivalent-information}
Consider two state-of-the-art distributed certificates, the scaled relative graph (SRG) and the $x$--$z$ graph. In \cite{Leng_Geometric_2026}, a one-to-one mapping between the upper-half SRG and the $x$--$z$ graph is established for complex matrices. 
Suppose the system dynamics are described by the transfer matrix $G(s)$.
At a fixed frequency, let
\begin{align}
\Ical_{xz}(G)&:=P_{xz}(G),\\
\Ical_{\rm SRG}(G)&:=\operatorname{SRG}_{+}(G).
\end{align}
The two representations are related by the bijections $(x,z)\mapsto x+\mathrm{j}\sqrt{z-x^2}$ and $
c\mapsto \bigl(\Re(c),|c|^2\bigr)$ \cite{Leng_Geometric_2026}.
The mapping is a bijection and thus can recover either report deterministically from the other. Hence
\begin{equation}
\Ical_{xz}\succeq \Ical_{\rm SRG},
\qquad
\Ical_{\rm SRG}\succeq \Ical_{xz},
\end{equation}
so the two representations induce exactly the same information fibers and therefore the same information-limited optimum:
\begin{equation}
\Fcal_{\Ical_{xz}}(\Sigma)=\Fcal_{\Ical_{\rm SRG}}(\Sigma),
\qquad
c_{\Ical_{xz}}^\star(\Sigma)=c_{\Ical_{\rm SRG}}^\star(\Sigma).
\end{equation}
Consequently, they also have identical information-induced conservatism. If an SRG-based test and an $x$--$z$-graph-based test exhibit different conservatism, the difference cannot come from the information structure. It must arise from how the same information is exploited, for example through an over-approximation or a restricted separation procedure. In other words,
\begin{equation}
    \Delta_{\rm info}(\Sigma; \Ical_{xz}) = \Delta_{\rm info}(\Sigma; \Ical_{\rm SRG})
\end{equation}
In the terminology of \eqref{eq:conservatism-decomposition}, the difference is method-induced rather than information-induced.
\end{example}

\begin{example}[Information Order of Geometric Distributed Certificates]
\label{ex:geometric-information-order}
The geometric constructions underlying several decentralized certificates \cite{Huang_ProjectingDW_2025,Leng_Geometric_2026} can be interpreted as an information partial order, rather than as a single chain of increasingly sophisticated methods. For a matrix $G$ at a fixed frequency, define the Davis--Wielandt-shell (DW-shell) $\Ical_{\rm DW}(G)$, numerical-range $\Ical_{W}(G)$, and $x$--$z$ reports $\Ical_{xz}(G)$ by
\begin{align}
\Ical_{\rm DW}(G)&:=DW(G),\\
\Ical_{W}(G)&:=W(G),\\
\Ical_{xz}(G)&:=P_{xz}(G).
\end{align}
We refer to \cite{Huang_Decentralized_2026} for a detailed and latest tutorial of these theories.
The numerical range and the $x$--$z$ graph are deterministic projections of the DW shell \cite{Huang_ProjectingDW_2025}. Moreover, the matrix phase bounds are determined from the supporting rays of the numerical range \cite{Chen_Phase_2024}, while a gain report such as $\|G\|_2$ can be recovered from $P_{xz}(G)$ through $\|G\|_2^2=\max_{(x,z)\in P_{xz}(G)}z$. Together with the bijection between $P_{xz}(G)$ and $\operatorname{SRG}_{+}(G)$ \cite{Leng_Geometric_2026}, this gives
\begin{equation}
\Ical_{\rm DW}\succeq\Ical_W\succeq\Ical_{\rm phase},
\qquad
\Ical_{\rm DW}\succeq\Ical_{xz}\equiv\Ical_{\rm SRG}\succeq\Ical_{\rm gain}.
\label{eq:geometry-information-partial-order}
\end{equation}
Accordingly, \eqref{eq:cert-order} predicts that the information-limited optimum cannot deteriorate when moving upward along either branch. However, $W(G)$ and $P_{xz}(G)$ are generally incomparable: the former retains imaginary/phase geometry that the latter removes, whereas the latter retains norm information that the former does not determine. Thus a universal conservatism ordering between numerical-range and $x$--$z$/SRG certificates cannot be inferred from information alone.

Integral quadratic constraints (IQCs) require a separate distinction. Frequency-domain quadratic-constraint formulations provide a flexible certificate framework through a chosen family of multipliers \cite{Liu_Scalable_2026}. An IQC is therefore not, by itself, a unique information report: its effective information structure depends on which input--output quantities are exposed, while the multiplier family determines how those quantities are used. Enlarging a multiplier family with the same report can reduce method-induced conservatism without changing $\Delta_{\rm info}$. Conversely, exposing richer quadratic/geometric data refines the information structure itself. 
\end{example}

If a refinement augments $\Ical$ by additional information $\Jcal$, then
$c^\star_{\{\Ical,\Jcal\}}(\Sigma)\ge c^\star_{\Ical}(\Sigma)$.

\begin{definition}[Stability Value of Information]
\label{def:value-information}
For additional information $\Jcal$, define its local value relative to $\Ical$
at $\Sigma$ as
\begin{equation}
\Delta c^*(\Jcal\mid\Ical;\Sigma)
:=
c_{\{\Ical,\Jcal\}}^{\star}(\Sigma)
-
c_{\Ical}^{\star}(\Sigma).
\label{eq:value-information}
\end{equation}
\end{definition}

\begin{proposition}
For additional information $\Jcal$, the local value relative to $\Ical$ at $\Sigma$ satisfies
    \begin{equation}
        \Delta c^*(\Jcal\mid\Ical;\Sigma)=-(\Delta_{\rm info}(\Sigma;\{\Ical,\Jcal\})-(\Delta_{\rm info}(\Sigma;\Ical))
    \end{equation}
\end{proposition}
\begin{proof}
    It is straightforward to substitute \eqref{eq:value-information} into \eqref{eq:information-gap}.
\end{proof}

The definition measures information in the same unit as the stability task.
A high-dimensional report can have zero value if it removes no system that
determines the fiber worst case, while one directional quantity can have
 substantial value if it separates the critical stable and unstable systems. 
\emph{In other words, the amount of reported information need not indicate its value
for a specific stability-evaluation task.}

\section{Information--Conservatism Frontier from the Perspective of Information Equivalence Classes}
\label{sec:equivalence-limits}

\subsection{Information-Induced Equivalence Classes}

 The information structure in Definition~\ref{def:information-structure} does more 
than reduce the dimension of the available description: it partitions the
physical system class according to what the certificate can and cannot
distinguish. In other words, it leads to classes of systems that are indistinguishable under the information structure.

\begin{proposition}[Information-Induced Equivalence Relation and Quotient Space]
\label{cor:information-quotient}
For a fixed information structure $\Ical:\mathcal C\rightarrow\mathcal Z$, define
the relation $\sim_{\Ical}$ on $\mathcal C$ as
\begin{equation}
\Sigma_1\sim_{\Ical}\Sigma_2
\quad\Longleftrightarrow\quad
\Ical(\Sigma_1)=\Ical(\Sigma_2).
\label{eq:information-equivalence}
\end{equation}
Then $\sim_{\Ical}$ is an equivalence relation, and the corresponding quotient
space
\begin{equation}
\mathcal C/{\sim_{\Ical}}
=
\left\{[\Sigma]_{\Ical}:\Sigma\in\mathcal C\right\}
\label{eq:information-quotient-space}
\end{equation}
is the space of information-induced indistinguishable system classes. Each
equivalence class satisfies
 $[\Sigma]_{\Ical}=\Fcal_{\Ical}(\Sigma)$. 
\end{proposition}

\begin{proof}
Reflexivity follows from $\Ical(\Sigma)=\Ical(\Sigma)$. Symmetry follows
because $\Ical(\Sigma_1)=\Ical(\Sigma_2)$ implies
$\Ical(\Sigma_2)=\Ical(\Sigma_1)$. Transitivity follows because
$\Ical(\Sigma_1)=\Ical(\Sigma_2)$ and
$\Ical(\Sigma_2)=\Ical(\Sigma_3)$ imply
$\Ical(\Sigma_1)=\Ical(\Sigma_3)$. Hence $\sim_{\Ical}$ is an equivalence
relation. By definition, $[\Sigma]_{\Ical}$ contains exactly the systems with
the same information representation as $\Sigma$, which is
 $\Fcal_{\Ical}(\Sigma)$. 
\end{proof}

Consequently, a method restricted to $\Ical$ operates on
$\mathcal C/{\sim_{\Ical}}$, rather than on an individual physical realization.
The stability task is therefore preserved exactly only when the information
partition is sufficiently fine for that task.

\subsection{Task Sufficiency and Within-Class Conservatism}

 By Theorem~\ref{thm:optimal-certificate}, the pointwise information-induced 
conservatism can be written directly on the equivalence class as
\begin{equation}
\boxed{
\Delta_{\rm info}(\Sigma;\Ical)
=
m(\Sigma)
-
 \inf_{\Sigma'\in\Fcal_{\Ical}(\Sigma)} 
m(\Sigma')}.
\label{eq:exact-gap}
\end{equation}
Thus, once the information structure is fixed, the irreducible conservatism is
completely determined by the variation of the stability task within the
 corresponding equivalence class. 

 \begin{theorem}[Equivalence-Class Width and Information-Induced Conservatism] 
\label{thm:class-width}
 For any reported value $z\in\Ical(\mathcal C)$, 
\begin{equation}
\sup_{\Sigma\in\Fcal_{\Ical}(z)}
\Delta_{\rm info}(\Sigma;\Ical)
=
\sup_{\Sigma\in\Fcal_{\Ical}(z)}m(\Sigma)
-
\inf_{\Sigma\in\Fcal_{\Ical}(z)}m(\Sigma).
\label{eq:class-width}
\end{equation}
 Consequently, the worst-case information-induced conservatism of the information 
structure is
\begin{equation}
\begin{aligned}
\overline\Delta_{\rm info}(\Ical)
:=
\sup_{\Sigma\in\mathcal C}\Delta_{\rm info}(\Sigma;\Ical)=
\sup_{z\in\Ical(\mathcal C)}
\left[
\sup_{\Sigma\in\Fcal_{\Ical}(z)}m(\Sigma)
-
\inf_{\Sigma\in\Fcal_{\Ical}(z)}m(\Sigma)
\right].
\end{aligned}
\label{eq:fiber-spread}
\end{equation}
\end{theorem}

\begin{proof}
For fixed $z$, Theorem~\ref{thm:optimal-certificate} gives the same value
$c_{\Ical}^{\star}(z)=\inf_{\Sigma'\in\Fcal_{\Ical}(z)}m(\Sigma')$ for every
system in $\Fcal_{\Ical}(z)$. Taking the supremum of
$m(\Sigma)-c_{\Ical}^{\star}(z)$ over that class proves
\eqref{eq:class-width}. Taking the supremum over all nonempty information
classes gives \eqref{eq:fiber-spread}.
\end{proof}

Theorem~\ref{thm:class-width} shows that the relevant ``size'' of an
 equivalence class is not its geometric diameter in the full system 
space, but its width along the stability task $m$. A large unresolved set can
 have negligible information-induced conservatism if $m$ varies little over it, 
whereas a narrow class can still be costly when it crosses a steep
stability-relevant direction.

The zero-width case gives the exact condition under which the stability margin
itself can be evaluated on the quotient space.

\begin{theorem}[Exact Certificate on the Information Quotient]
\label{thm:quotient-exact-certificate}
Let $q_{\Ical}:\mathcal C\rightarrow\mathcal C/{\sim_{\Ical}}$ denote the natural
projection $q_{\Ical}(\Sigma)=[\Sigma]_{\Ical}$. The following statements are
equivalent:
\begin{enumerate}
\item the information structure introduces no irreducible conservatism,
i.e., $\Delta_{\rm info}(\Sigma;\Ical)=0$ for all $\Sigma\in\mathcal C$;
\item the ground-truth margin is constant on every information fiber, i.e.,
\begin{equation}
\Sigma_1\sim_{\Ical}\Sigma_2
\quad\Longrightarrow\quad
m(\Sigma_1)=m(\Sigma_2);
\label{eq:margin-constant-on-fiber}
\end{equation}
\item there exists a unique map
$\bar m_{\Ical}:\mathcal C/{\sim_{\Ical}}\rightarrow\R$ such that
\begin{equation}
m=\bar m_{\Ical}\circ q_{\Ical}.
\label{eq:margin-factor-quotient}
\end{equation}
\end{enumerate}
Hence exact stability-margin evaluation from the compressed information is
possible if and only if the stability task itself is well defined on the
information-induced quotient space.
\end{theorem}

\begin{proof}
If $\Delta_{\rm info}(\Sigma;\Ical)=0$ for every $\Sigma$, then
\eqref{eq:information-gap} gives
$c_{\Ical}^{\star}(\Ical(\Sigma))=m(\Sigma)$. Therefore, any two systems in the
same information fiber have the same value of $m$, which proves
\eqref{eq:margin-constant-on-fiber}. Conversely, if $m$ is constant on every
fiber, then \eqref{eq:optimal-certificate} gives
$c_{\Ical}^{\star}(\Ical(\Sigma))=m(\Sigma)$ and thus
$\Delta_{\rm info}(\Sigma;\Ical)=0$.
Finally, a function on $\mathcal C$ is constant on the equivalence classes of
$\sim_{\Ical}$ if and only if it factors uniquely through the quotient map
$q_{\Ical}$, which establishes the equivalence with
\eqref{eq:margin-factor-quotient}.
\end{proof}

Equation~\eqref{eq:margin-factor-quotient} means that an exact certificate is
possible precisely when the ground-truth margin does not distinguish any two
systems that the information structure has already declared indistinguishable.
If this factorization fails, no refinement of the certificate map $c$ alone
can eliminate the resulting conservatism; the information structure itself
must be refined.

The same quotient-space result immediately contains the binary stability
problem as a special case.

\begin{example}[Binary Stability as Quotient Separation]
For binary stability certification, define
\begin{equation}
\begin{aligned}
\mathcal C_{\rm s}&:=\{\Sigma\in\mathcal C:m(\Sigma)>0\},\\
\mathcal C_{\rm u}&:=\{\Sigma\in\mathcal C:m(\Sigma)\le0\}.
\end{aligned}
\label{eq:stable-unstable-sets}
\end{equation}
Unlike exact margin evaluation, a binary decision does not require $m$ to be
constant on each equivalence class. It only requires the information
compression to preserve the stability boundary:
\begin{equation}
\boxed{q_{\Ical}(\mathcal C_{\rm s})\cap q_{\Ical}(\mathcal C_{\rm u})=\varnothing}.
\label{eq:quotient-stability-separation}
\end{equation}
 Equivalently, every equivalence class must lie entirely on one 
side of $m=0$. Hence binary stability is the natural sign-only consequence of
Theorem~\ref{thm:quotient-exact-certificate}, rather than a separate
information concept.
\end{example}

\begin{corollary}[Indistinguishable Stable/Unstable Pair]
\label{cor:indistinguishable-no-go}
Suppose that there exist $\Sigma_{\rm s},\Sigma_{\rm u}\in\mathcal C$ such that
\begin{equation}
\Ical(\Sigma_{\rm s})=\Ical(\Sigma_{\rm u}),
\qquad
m(\Sigma_{\rm s})>0,
\qquad
m(\Sigma_{\rm u})\le0.
\label{eq:stable-unstable-pair}
\end{equation}
Then no sound method using only $\Ical$ can certify
$\Sigma_{\rm s}$ as stable.
\end{corollary}

\begin{proof}
 Both systems belong to the same information fiber 
 $\Fcal_{\Ical}(\Sigma_{\rm s})=\Fcal_{\Ical}(\Sigma_{\rm u})$. 
Equation~\eqref{eq:optimal-certificate} gives
$c_{\Ical}^{\star}\le m(\Sigma_{\rm u})\le0$ on that class.
\end{proof}

\emph{Corollary~\ref{cor:indistinguishable-no-go} is the basic information wall:
the limitation is not that a particular distributed test fails, but that the
available information does not distinguish the stable system from a physically
admissible unstable one.} It also converts the quotient-space separation
condition above into a constructive impossibility test. Rather than
characterizing the entire quotient space, it is sufficient to exhibit a
 single stable/unstable pair that collapses to the same equivalence class. 
Such a pair is therefore a witness that the current information structure is
fundamentally insufficient, irrespective of the stability criterion used.
This pairwise construction will be used in
Section~\ref{sec:network-spectrum-insufficiency}.

\subsection{Information--Conservatism Frontier under Admissible Refinements}
\label{subsec:general-frontier}

The preceding results evaluate the limitation created by a \emph{fixed} information structure. To turn this benchmark into a design 
problem of reporting interfaces, the admissible ways of enriching the report must first be specified.
This restriction is essential. If arbitrary information structures were allowed, the single real-valued report $\Ical(\Sigma)=m(\Sigma)$ would make the information-induced conservatism identically zero. Consequently, the raw dimension of the report cannot define a meaningful information budget by itself.
This is the significant distinction: more information in the Shannon sense does not necessarily bring more value for stability assessment.

Fix a baseline reporting interface with information structure $\Ical_0$ and a finite family of admissible information primitives
\begin{equation}
\Lambda=\{\Jcal_1,\ldots,\Jcal_p\}.
\label{eq:primitive-family}
\end{equation}
For a selected index set $A\subseteq[p]:=\{1,\ldots,p\}$, define the refined
 information structure 
\begin{equation}
\Ical_A
:=
\{\Ical_0,\Jcal_\ell:\ell\in A\},
\qquad
 \operatorname{cost}(A):=\sum_{\ell\in A}w_\ell, 
\quad w_\ell>0,
\label{eq:admissible-refinement}
\end{equation}
where $w_\ell$ is the cost assigned to the $\ell$-th reporting primitive.
In practical systems, $w_\ell$ may represent the cost of additional sensing and communication resources, such as PMU or SCADA measurements.
In the sense of information, the cost may represent one additional directional report, one certified cross-node relation, one model block, or another reporting-interface resource.

\begin{definition}[Information--Conservatism Frontier]
\label{def:information-frontier}
For an information budget $B\ge0$, define the worst-case frontier
\begin{equation}
\boxed{
\Phi(B)
:=
 \min_{A\subseteq[p]:\,\operatorname{cost}(A)\le B} 
\overline\Delta_{\rm info}(\Ical_A)}.
\label{eq:information-frontier}
\end{equation}
For a fixed physical system $\Sigma$, its pointwise counterpart is
\begin{equation}
\Phi(B;\Sigma)
:=
 \min_{A\subseteq[p]:\,\operatorname{cost}(A)\le B} 
\Delta_{\rm info}(\Sigma;\Ical_A).
\label{eq:point-wise-information-frontier}
\end{equation}
Conversely, for a prescribed tolerance $\varepsilon\ge0$, define the minimum
information cost
\begin{equation}
\boxed{
 \operatorname{cost}^\star(\varepsilon) 
:=
\min_{A\subseteq[p]}
 \left\{\operatorname{cost}(A): 
\overline\Delta_{\rm info}(\Ical_A)\le\varepsilon
\right\},}
\label{eq:minimum-information-cost}
\end{equation}
 with $\operatorname{cost}^\star(\varepsilon)=+\infty$ when the admissible refinement family 
cannot attain the requested tolerance.
\end{definition}

By Theorem~\ref{thm:class-width}, the frontier has the equivalent
partition-width form
\begin{equation}
\Phi(B)
=
 \min_{A:\,\operatorname{cost}(A)\le B} 
\sup_{z\in\Ical_A(\mathcal C)}
\left[
\sup_{\Sigma\in\Fcal_{\Ical_A}(z)}m(\Sigma)
-
\inf_{\Sigma\in\Fcal_{\Ical_A}(z)}m(\Sigma)
\right].
\label{eq:frontier-partition-width}
\end{equation}
Hence the information-design problem is geometrically a partition-design
 problem: under a prescribed reporting budget, refine the information 
 fibers so that their largest width along the task $m$ is as small as 
possible. The frontier is nonincreasing in $B$, and the two formulations obey
\begin{equation}
\Phi(B)\le\varepsilon
\quad\Longleftrightarrow\quad
 \operatorname{cost}^\star(\varepsilon)\le B. 
\label{eq:frontier-inverse-relation}
\end{equation}
 In particular, $\operatorname{cost}^\star(0)$ is the minimum admissible reporting-interface cost required 
for exact task sufficiency.

The minimum-information problem also admits a pairwise characterization that
extends Corollary~\ref{cor:indistinguishable-no-go}. For $\varepsilon\ge0$,
define the baseline task-relevant witness set
\begin{equation}
\begin{aligned}
\mathcal W_\varepsilon
:=\big\{(\Sigma_a,\Sigma_b)\in\mathcal C^2:\;&
\Ical_0(\Sigma_a)=\Ical_0(\Sigma_b),\\
&|m(\Sigma_a)-m(\Sigma_b)|>\varepsilon\big\}.
\end{aligned}
\label{eq:witness-set-epsilon}
\end{equation}
and, for each pair, the primitives capable of separating it,
\begin{equation}
D(\Sigma_a,\Sigma_b)
:=
\left\{\ell\in[p]:
\Jcal_\ell(\Sigma_a)\ne\Jcal_\ell(\Sigma_b)
\right\}.
\label{eq:witness-separating-primitives}
\end{equation}

\begin{theorem}[Task-Relevant Witness Cover]
\label{thm:witness-cover}
For any selected refinement $A\subseteq[p]$ and any $\varepsilon\ge0$,
\begin{equation}
\boxed{
\begin{aligned}
\overline\Delta_{\rm info}(\Ical_A)\le\varepsilon
\quad\Longleftrightarrow\quad
A\cap D(\Sigma_a,\Sigma_b)\ne\varnothing,\;\forall(\Sigma_a,\Sigma_b)\in\mathcal W_\varepsilon.
\end{aligned}}
\label{eq:witness-cover-equivalence}
\end{equation}
\end{theorem}

\begin{proof}
Suppose first that a pair
$(\Sigma_a,\Sigma_b)\in\mathcal W_\varepsilon$ is not separated by any
primitive in $A$. Since the two systems already have the same baseline report,
they also satisfy $\Ical_A(\Sigma_a)=\Ical_A(\Sigma_b)$ and therefore belong
to the same $\Ical_A$-fiber. The width of that fiber along $m$ is strictly
larger than $\varepsilon$, so Theorem~\ref{thm:class-width} gives
$\overline\Delta_{\rm info}(\Ical_A)>\varepsilon$.

Conversely, suppose
$\overline\Delta_{\rm info}(\Ical_A)>\varepsilon$. By
\eqref{eq:fiber-spread}, some $\Ical_A$-fiber has $m$-width greater than
$\varepsilon$. Hence it contains two systems $\Sigma_a,\Sigma_b$ satisfying
$|m(\Sigma_a)-m(\Sigma_b)|>\varepsilon$. They have the same baseline report
and are not distinguished by any selected primitive, so
$(\Sigma_a,\Sigma_b)\in\mathcal W_\varepsilon$ and
$A\cap D(\Sigma_a,\Sigma_b)=\varnothing$.
\end{proof}

Theorem~\ref{thm:witness-cover} makes the frontier constructive: additional
information has value only through the task-relevant pairs that it separates.
For a finite admissible system set, \eqref{eq:minimum-information-cost} is
exactly the weighted covering problem
\begin{equation}
\boxed{
\begin{aligned}
\min_{x_1,\ldots,x_p}\quad
&\sum_{\ell=1}^{p}w_\ell x_\ell\\
\mathrm{s.t.}\quad
&\sum_{\ell\in D(\Sigma_a,\Sigma_b)}x_\ell\ge1,
&&\forall(\Sigma_a,\Sigma_b)\in\mathcal W_\varepsilon,\\
&x_\ell\in\{0,1\},
&&\ell=1,\ldots,p.
\end{aligned}
}
\label{eq:witness-cover-mip}
\end{equation}
For a continuous system class, the problem becomes a semi-infinite covering problem. This suggests a constraint-generation procedure: refine the reporting interface only when an indistinguishable pair with task separation greater than $\varepsilon$ is found. For binary
stability certification, $\mathcal W_\varepsilon$ is replaced by the
 stable/unstable witness pairs of Corollary~\ref{cor:indistinguishable-no-go}; 
thus minimum sufficient information is precisely the minimum-cost refinement
that separates every such pair.

 \section{Network-Node Feedback Interconnection Structure and a Unified View of Information-Induced Conservatism} 
\label{sec:distributed-structure}

Distributed stability methods exploit the modular structure of an interconnected system. A common physical representation separates local node dynamics from network interconnections:

\begin{equation}
\begin{aligned}
\text{node: }& y=G\circ u,\; G=\operatorname{blkdiag}\!\bigl(G_1,\ldots,G_n\bigr),\\
\text{network: }& u=-N\circ y,\; 
\end{aligned}
\label{eq:network-node-generic}
\end{equation}
Here $G_i$ is the dynamics of node $i$ and $N$ describes the network interconnection. This representation is widely used in practical system modeling such as power grids \cite{Haberle_Decentralized_2026, Peng_Compositional_2026}. Different port selections lead to different realizations, but the architectural principle is the same: local dynamics and network interactions are separated so that certification can be assembled modularly.

\begin{figure*}[htbp]
\centering
\begin{tikzpicture}[>=stealth, every node/.style={font=\scriptsize,text=black}]
    \tikzset{block/.style={draw=black,rounded corners,align=center,minimum height=0.85cm,inner sep=4pt}}
    \node[block,text width=3.8cm] (nodes) at (0,2.1) {Node dynamics\\$G=\operatorname{blkdiag}(G_1,\ldots,G_n)$};
    \node[block,text width=3.8cm] (network) at (7.2,2.1) {Network interconnection dynamics\\$N$};
    \draw[->,thick,black] (nodes.east) -- node[above] {$y=G\circ u$} (network.west);
    \draw[->,thick,black] (network.south) .. controls (6.2,1.0) and (1.0,1.0) .. node[above] {$u=-N\circ y$} (nodes.south);

    \node[block,text width=2.45cm] (inode) at (0,0) {Node report\\$\Ical_{\rm node}$};
    \node[block,text width=2.65cm] (ijoint) at (3.6,0) {Relational report\\$\Ical_{\rm joint}$};
    \node[block,text width=2.45cm] (inet) at (7.2,0) {Network report\\$\Ical_{\rm net}$};
     \node[block,text width=3.2cm] (cert) at (3.6,-1.5) {Distributed certificate\\uses only the reporting interface}; 

    \draw[->,thick,black] (nodes.south) -- (inode.north);
    \draw[->,thick,black] (network.south) -- (inet.north);
    \draw[->,dashed,black] (nodes.south east) -- (ijoint.north west);
    \draw[->,dashed,black] (network.south west) -- (ijoint.north east);
    \draw[->,thick,black] (inode.south) -- (cert.north west);
    \draw[->,thick,black] (ijoint.south) -- (cert.north);
    \draw[->,thick,black] (inet.south) -- (cert.north east);
\end{tikzpicture}
\caption{{Physical decomposition and reporting-interface decomposition in distributed stability certification. The network--node model specifies how the physical system is assembled; the reporting interface induces the information structure that specifies what the certificate can distinguish. The three studies below correspond to compression of the network report, compression of the node report, and omission of relational information, respectively.}}
\label{fig:distributed-node-network-split}
\end{figure*}
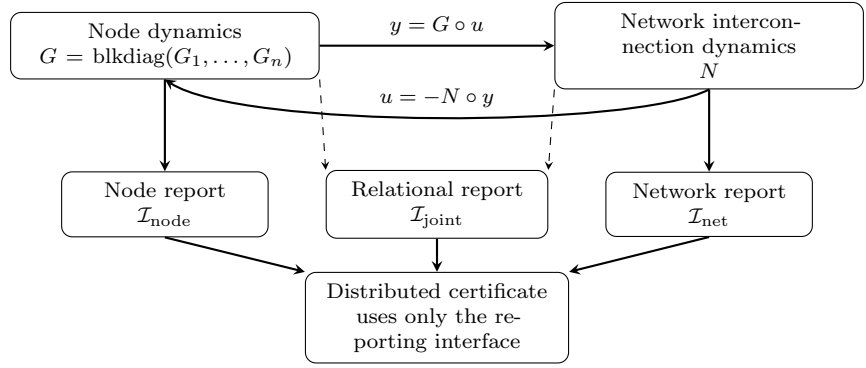

For the present paper, the crucial point is that the reporting interface induces an information structure of the form 
\[
\Ical(\Sigma)=\bigl\{\Ical_{\rm node},\ \Ical_{\rm net},\ \Ical_{\rm joint}\bigr\},
\]
where the first two terms represent the node and network reports and $\Ical_{\rm joint}$ represents any retained relation between them or across nodes. A practical reporting interface may compress either side and may omit $\Ical_{\rm joint}$ entirely. Distributed certification is therefore not only a problem of exploiting a physical decomposition; it is also a problem of designing the reporting interface that determines what the certificate is allowed to observe.

The physical decomposition does not, by itself, determine what the certificate knows. The three application studies therefore vary the \emph{reporting interface}, not the underlying idea of modular interconnection. Table~\ref{tab:three-information-losses} places the three sources of information-induced conservatism side by side and makes explicit the relation that each reporting interface fails to preserve.

The three cases expose a common structural message: stability can depend on relations among quantities reported by all interfaces. Network-side compression loses the placement of network directions relative to heterogeneous nodes. Node-side compression loses local eigendirections relative to the critical interconnection direction. Cross-node compression loses relations that do not belong to any single local block. In every case, the information structure places systems with potentially different stability margins in the same information fiber. Consequently, the stability value of information is governed by how strongly the task varies along the unresolved relation, rather than by the number of scalars added to or removed from the report. 

\begingroup 
\begin{table*}[htpb]
 
\centering
\caption{Three representative sources of information-induced conservatism in a modular network--node architecture. The common issue is not simply that quantities are omitted, but that stability-relevant relations become unresolved.}
\label{tab:three-information-losses}
\begin{tabularx}{\textwidth}{@{}p{0.14\textwidth}p{0.27\textwidth}X p{0.11\textwidth}@{}}
\toprule
 Source of conservatism & What remains available & Stability-relevant relation left unresolved & Studied in \\
\midrule
Network-side &
Complete node models and the network Laplacian eigenvalues &
Alignment between network eigendirections and heterogeneous node characteristics &
Section~\ref{sec:network-spectrum-insufficiency}\\
Node-side &
Complete network certificate information and node scalar or eigenvalue-only passivity reports &
Alignment between node passivity eigendirections and the critical interconnection direction &
Section~\ref{sec:passivity-information-value}\\
Cross-node &
Exact local matrix blocks assembled from independently certified nodes &
Cross-node quadratic relations that cannot be reconstructed from the local blocks alone &
Section~\ref{sec:cross-node-info-value}\\
\bottomrule
\end{tabularx}
\end{table*}
\endgroup

 This interpretation leads to a concrete diagnostic workflow. \emph{Step 1:} specify the stability task and its margin. \emph{Step 2:} write the information structure induced by the reporting interface, including what joint relations are unavailable. \emph{Step 3:} characterize the adverse variation inside the resulting information fiber, either by computing $c_{\Ical}^{\star}$ or by constructing an indistinguishable witness pair. \emph{Step 4:} compare the two gaps in \eqref{eq:conservatism-decomposition}. 
Large method-induced conservatism $\Delta_{\rm method}$ calls for a better criterion or optimization. By contrast, large information-induced conservatism $\Delta_{\rm info}$ calls for a richer reporting interface. If both terms are small, the existing reporting interface is already close to the task-relevant limit. The following sections apply this workflow to the three reporting interfaces in Table~\ref{tab:three-information-losses}.

 \section{Network-Side Information-Induced Conservatism under Spectral Compression} 
\label{sec:network-spectrum-insufficiency}

The preceding results characterize the best stability certificate permitted by
 a given information structure. We now examine a common source of information-induced 
 conservatism on the network side of distributed assessment: replacing the network 
matrix by its eigenvalue spectrum. To isolate this compression effect, the node
side is assumed to be known \emph{completely}. The question is whether network
eigenvalues alone, without the associated eigenvectors or their alignment with
heterogeneous nodes, are sufficient for exact stability certificates.

\subsection{Insufficiency of Network Spectrum Information}

Consider the linearized system dynamics $y=G(s)u$, $u=-bLy$, where $L=L^T\succeq0$ is the Laplacian of a connected network, $b>0$ is a strength scaling, and the complete node dynamics are
\begin{equation}
G(s)=g_0(s)K,\qquad
K=\diag(k_1,\ldots,k_n)\succ0.
\label{eq:app1-common-shape}
\end{equation}
Although \eqref{eq:app1-common-shape} is a restricted subclass of heterogeneous
nodes, an impossibility result on this subclass is already sufficient to
disprove universal sufficiency of network-spectrum information for heterogeneous systems.

Since $KL$ is similar to $K^{1/2}LK^{1/2}$, the closed-loop characteristic
determinant factorizes exactly as
\begin{equation}
\begin{aligned}
\det[I+bG(s)L]=\det[I+bg_0(s)K^{1/2}LK^{1/2}]=\prod_{j=1}^{n}\left[1+b\mu_j g_0(s)\right],\;\hspace{0.8cm}\mu_j\in\spec(K^{1/2}LK^{1/2}).
\end{aligned}
\label{eq:app1-matched-factorization}
\end{equation}
Thus, once the nodes are heterogeneous, the stability-relevant network
quantity is generally the matched spectrum
$\spec(K^{1/2}LK^{1/2})$, rather than $\spec(L)$ alone.
A stronger statement follows.

\begin{proposition}[Network Spectrum Is Not Universally Sufficient under Heterogeneity]
\label{thm:network-spectrum-insufficient}
Fix any $n\ge4$ and a stable, real-rational, strictly proper common node shape
 $g_0(s)=n(s)/d(s)$. Suppose that the scalar modal family 
\begin{equation}
 \chi_\mu(s):=d(s)+\mu n(s),\qquad \mu\ge0, 
\label{eq:app1-general-modal-polynomial}
\end{equation}
has a finite first positive-gain loss-of-Hurwitz boundary
\begin{equation}
0<\mu_c:=
\inf\{\mu>0:\chi_\mu(s)\ \text{is not Hurwitz}\}<\infty .
\label{eq:app1-general-muc}
\end{equation}
Then there exist a heterogeneous gain matrix $K\succ0$ and two connected
$n$-node Laplacians $L_A,L_B$ such that
\[
\spec(L_A)=\spec(L_B),
\]
 but one closed-loop interconnection is asymptotically stable while the other is {not Hurwitz}. Consequently, the information structure 
\begin{equation}
\Ical_{\rm spec}^{L}(\Sigma)
:=
\{G_1(s),\ldots,G_n(s),\spec(L)\}
\label{eq:app1-spectrum-information}
\end{equation}
is not universally sufficient for exact stability evaluation on this
heterogeneous system class. A sufficient condition for the finite-boundary
assumption above is that $g_0$ has relative degree at least three.
\end{proposition}

\begin{proof}
In the proof, we consider the star topology introduced in Section~\ref{sec:introduction}.
Because $d$ is Hurwitz, $\chi_\mu$ remains Hurwitz for all sufficiently small  $\mu\ge0$, so $\mu_c>0$. The proposition assumes that a finite loss-of-Hurwitz boundary exists. 
The commonly encountered relative-degree condition $r:=\deg d-\deg n\ge3$ guarantees this finiteness. The root locus of $d(s)+\mu n(s)$ then has $r$ branches tending to infinity, and at least one asymptote enters the open right-half plane in the high-frequency bands. Hence $\chi_\mu$ is non-Hurwitz for some finite positive
$\mu$.

It remains to construct two networks with distinct stability and the same
spectrum. Let
\[
K=\diag(\kappa,1,\ldots,1),\qquad \kappa>1.
\]
Consider an unweighted $n$-node star. In Network A, node $1$, carrying the gain
$\kappa$, is the center. In Network B, node $2$ is the center, so the same
labeled heterogeneous node $1$ becomes a leaf. The two Laplacians have exactly
the same complete spectrum,
\begin{equation}
\spec(L_A)=\spec(L_B)=\{0,1,\ldots,1,n\}.
\label{eq:app1-star-spectrum}
\end{equation}
The node models remain fixed at their labeled nodes and are therefore identical
in the two systems.

For Network A,
\begin{equation}
\lambda_{L,A}:=\lambda_{\max}(K^{1/2}L_AK^{1/2})
=\kappa(n-1)+1.
\end{equation}
For Network B, the $n-3$ directions among the ordinary leaves have eigenvalue
$1$, while the remaining nonzero matched eigenvalues are the roots of
$p(\lambda)=\lambda^2-(n+\kappa)\lambda+\kappa(n-1)+1$. Hence
\begin{equation}
\lambda_{L,B}
=
\frac{
n+\kappa+
\sqrt{(n+\kappa)^2-4[\kappa(n-1)+1]}
}{2}.
\label{eq:app1-star-leaf}
\end{equation}
Substituting $\lambda_{L,A}$ gives
\begin{equation}
p(\lambda_{L,A})
=
(\kappa-1)(n-2)[\kappa(n-1)+1]>0.
\end{equation}
Moreover, for $n\ge4$ and $\kappa>1$,
$\lambda_{L,A}>(n+\kappa)/2$, so $\lambda_{L,A}$ lies to the right of the
vertex of $p$. Therefore
\begin{equation}
\lambda_{L,A}>\lambda_{L,B}.
\label{eq:app1-star-gap}
\end{equation}

Choose
\[
b=\frac{\mu_c}{\lambda_{L,A}}.
\]
For Network B, every matched eigenvalue is nonnegative and satisfies
\begin{equation}
b\mu_j
\le
b\lambda_{L,B}
=
\mu_c\frac{\lambda_{L,B}}{\lambda_{L,A}}
<\mu_c.
\end{equation}
By the definition of the first boundary, every scalar factor in
\eqref{eq:app1-matched-factorization} is Hurwitz, and Network B is
asymptotically stable. In Network A, the largest matched mode satisfies
$b\lambda_{L,A}=\mu_c$. Since the set of Hurwitz polynomials is open and the
coefficient map is continuous, the boundary polynomial $\chi_{\mu_c}$ is
non-Hurwitz. Hence Network A is not Hurwitz. This produces an
 indistinguishable stable/unstable pair under 
$\Ical_{\rm spec}^{L}$ and proves the result.
\end{proof}

The proposition should be read as a statement about \emph{universal information sufficiency}, not as a claim that every heterogeneous node model produces an isospectral stability mismatch. A single admissible heterogeneous subclass containing such a pair is enough to show that the report $\{G_i,\spec(L)\}$ cannot support an exact certificate that is valid for the whole class. \emph{The missing quantity is the alignment between network modes and heterogeneous node characteristics.} This distinction disappears in the homogeneous limit $K=kI$, because then $K^{1/2}LK^{1/2}=kL$ and the matched spectrum is determined completely by $\spec(L)$. Thus the limitation is not caused by spectral analysis itself, but by compressing away a node--network relation that becomes stability-relevant under heterogeneity.

\subsection{Quantifying the Conservatism of Spectral Compression}
\label{subsec:network-spectrum-conservatism}

The preceding theorem proves that the network spectrum can be insufficient.
The information framework also allows the resulting conservatism to be
quantified. For the common-shape class in
\eqref{eq:app1-common-shape}, consider the margin
\begin{equation}
m(\Sigma)
=
\mu_c
-
b\lambda_{\max}(K^{1/2}LK^{1/2}),
\label{eq:network-first-boundary-margin}
\end{equation}
where $\mu_c$ is defined in \eqref{eq:app1-general-muc}.

If the full labeled network matrix $L$ is available, this margin is known exactly. If only the spectral report \eqref{eq:app1-spectrum-information} is available, the node matrix $K$ remains fixed while $L$ can vary over all admissible isospectral networks in the same information fiber. Therefore, Theorem~\ref{thm:optimal-certificate} gives

\begin{equation}
c_{\Ical_{\rm spec}^{L}}^\star(\Sigma)
=
\mu_c
-b
\sup_{\Sigma'\in\Fcal_{\Ical_{\rm spec}^{L}}(\Sigma)}
\lambda_{\max}
\!\left(K^{1/2}L'K^{1/2}\right).
\label{eq:network-spectrum-certificate}
\end{equation}
Consequently, the exact conservatism introduced by compressing the network
matrix into its spectrum is
\begin{equation}
\begin{aligned}
\Delta_{\rm info}(\Sigma;\Ical_{\rm spec}^{L})
=
b\Big[\sup_{\Sigma'\in\Fcal_{\Ical_{\rm spec}^{L}}(\Sigma)}
\lambda_{\max}(K^{1/2}L'K^{1/2})-\lambda_{\max}(K^{1/2}LK^{1/2})
\Big].
\end{aligned}
\label{eq:network-spectrum-exact-gap}
\end{equation}

Every node permutation $P^TLP$, while keeping the node matrix $K$ fixed at the
labeled nodes, is a physically admissible Laplacian with the same spectrum.
Hence
\begin{equation}
\begin{aligned}
\Delta_{\rm info}(\Sigma;\Ical_{\rm spec}^{L})
\ge b\Big[\max_{P\ {\rm permutation}}
\lambda_{\max}(K^{1/2}P^TLPK^{1/2})-\lambda_{\max}(K^{1/2}LK^{1/2})
\Big].
\end{aligned}
\label{eq:network-spectrum-permutation-bound}
\end{equation}
The lower bound isolates the conservatism caused purely by the unresolved
alignment of known nodes relative to a network with the same spectrum.

\begin{example}
The general result above can be seen in a four-node construction for which all
quantities are available in closed form. Fix the labeled node dynamics as
$G_i(s)=k_i/(s+1)^3$ with
$(k_1,k_2,k_3,k_4)=(\kappa,\kappa,1,1)$ and place them into the two networks
in Fig.~\ref{fig:app1-four-bus}. Network A places the two high-gain nodes on
adjacent nodes of the ring, whereas Network B places them on opposite nodes.

\begin{figure*}[htpb]
\centering
\begin{tikzpicture}[scale=0.89, every node/.style={font=\scriptsize}]
\node at (-2.45,1.35) {Network A: adjacent};
\coordinate (A1) at (-3.35,0.75);
\coordinate (A2) at (-1.55,0.75);
\coordinate (A3) at (-1.55,-0.85);
\coordinate (A4) at (-3.35,-0.85);
\draw[thick] (A1)--(A2)--(A3)--(A4)--cycle;
\foreach \p/\lab/\fillc in
{A1/1/gray!55,A2/2/gray!55,A3/3/white,A4/4/white}{
\filldraw[thick,fill=\fillc] (\p) circle (0.20);
\node at (\p) {$\lab$};
}

\node at (2.45,1.35) {Network B: opposite};
\coordinate (B1) at (1.55,0.75);
\coordinate (B3) at (3.35,0.75);
\coordinate (B2) at (3.35,-0.85);
\coordinate (B4) at (1.55,-0.85);
\draw[thick] (B1)--(B3)--(B2)--(B4)--cycle;
\foreach \p/\lab/\fillc in
{B1/1/gray!55,B2/2/gray!55,B3/3/white,B4/4/white}{
\filldraw[thick,fill=\fillc] (\p) circle (0.20);
\node at (\p) {$\lab$};
}
\node[anchor=west] at (-0.60,0.18) {gray: $k_i=\kappa$};
\node[anchor=west] at (-0.60,-0.20) {white: $k_i=1$};
\end{tikzpicture}
\caption{Four-node pair with identical labeled node dynamics and identical complete Laplacian spectrum. Only the alignment between node heterogeneity and network modes changes.}
\label{fig:app1-four-bus}
\end{figure*}
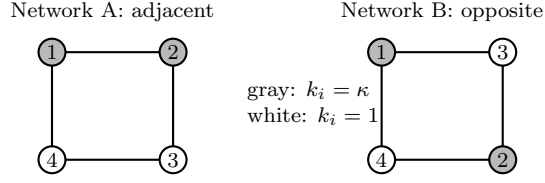

With unit line weights, the two Laplacians are
\begin{equation}
\begin{aligned}
L_A=
\begin{bmatrix}
2&-1&0&-1\\
-1&2&-1&0\\
0&-1&2&-1\\
-1&0&-1&2
\end{bmatrix},\;
L_B=
\begin{bmatrix}
2&0&-1&-1\\
0&2&-1&-1\\
-1&-1&2&0\\
-1&-1&0&2
\end{bmatrix}.
\end{aligned}
\label{eq:app1-four-laplacians}
\end{equation}
They are similar and satisfy
$\spec(L_A)=\spec(L_B)=\{0,2,2,4\}$. Nevertheless, for
$K=\diag(\kappa,\kappa,1,1)$, the largest matched eigenvalues are
\begin{equation}
\lambda_{L,A}=
\frac{3(\kappa+1)+\sqrt{9\kappa^2-14\kappa+9}}{2},
\qquad
\lambda_{L,B}=2(\kappa+1).
\label{equ:app1-four-eta}
\end{equation}
Their difference is strict for every $\kappa>1$ because
$9\kappa^2-14\kappa+9-(\kappa+1)^2=8(\kappa-1)^2>0$.
Take $\kappa=2.75$ and $b=1$. Then
$\lambda_{L,A}\approx8.72994$ and $\lambda_{L,B}=7.5$.
For $g_0(s)=1/(s+1)^3$, the Routh--Hurwitz criterion gives the first positive
boundary $\mu_c=8$. Hence Network A is unstable while Network B is
asymptotically stable.

 To quantify the resulting information-induced conservatism, take Network B as the true system. Its true 
first-boundary margin is $m(\Sigma_B)=8-\lambda_{L,B}=0.5$. The same spectral
fiber contains Network A, for which $\lambda_{L,A}\approx8.72994$. Thus
\begin{equation}
\Delta_{\rm info}(\Sigma_B;\Ical_{\rm spec}^{L})
\ge
\lambda_{L,A}-\lambda_{L,B}
\approx1.22994,
\label{eq:network-four-bus-gap}
\end{equation}
and consequently
$c_{\Ical_{\rm spec}^{L}}^\star(\Sigma_B)\le-0.72994$.
 Hence the spectrum-only reporting interface loses more stability margin than the true 
system possesses: although Network B is stable with margin $0.5$, no sound
certificate based only on the common spectrum can certify it.
\end{example}

 \section{Node-Side Information-Induced Conservatism under Spectral Compression}\label{sec:passivity-information-value}

The network-side result above shows that spectral compression can discard
stability-relevant eigenvector alignment. We next examine the complementary
 node-side information-induced conservatism caused by compressing matrix passivity indices into scalar 
reports. This section uses passivity-based certificates for demonstration, but
 the underlying mechanism is not specific to passivity. 

Consider the stacked passivity condition associated with
Example~\ref{ex:passivity-information},
\begin{equation}
\Pi-\Pi_g\succ0,
\qquad
\Pi=\operatorname{blkdiag}(\Pi_1,\ldots,\Pi_n),
\label{eq:app3-global-passivity}
\end{equation}
where $\Pi_i\in\mathbb S^2$ and $\Pi_g\in\mathbb S^{2n}$ denote the node and
network passivity indices, respectively. For this comparison, use the
passivity-residual margin
\begin{equation}
m(\Sigma)=\lambda_{\min}(\Pi-\Pi_g)
\label{eq:app3-passivity-margin}
\end{equation}
as the task margin. The full matrix report therefore provides the benchmark against the information-induced conservatism caused by scalarization.

\subsection{Passivity-Based Node Reports with Scalar, Spectral, and Matrix Values}

For each node, recall the scalar report
$\pi_i=\lambda_{\min}(\Pi_i)$. Since
$\Pi_i\in\mathbb S^2$, it can always be written as
\begin{equation}
\begin{aligned}
\Pi_i&=\pi_i I_2+r_i s_i s_i^T,\qquad \|s_i\|_2=1,\\
r_i&=\lambda_{\max}(\Pi_i)-\pi_i\ge0.
\end{aligned}
\label{eq:app3-Q-decomposition}
\end{equation}
The full eigenvalue spectrum is $\{\pi_i,\pi_i+r_i\}$.
 Hence, the scalar reporting interface retains only $\pi_i$, whereas a complete 
eigenvalue-spectrum report additionally reveals $\pi_i+r_i$ but still does not
reveal the direction $s_i$. Define
\begin{equation}
\Pi_{\rm sc}
:=
\operatorname{blkdiag}
(\pi_1 I_2,\ldots,\pi_n I_2).
\label{eq:app3-Qsc}
\end{equation}

To avoid confusing this node-side spectrum with the network spectrum used in Section~\ref{sec:network-spectrum-insufficiency}, denote by $\Ical_{\rm spec}^{\Pi}$ the refinement of $\Ical_{\rm sc}^{\Pi}$ that reports the complete eigenvalue spectrum of every $\Pi_i$, but not their eigenvectors.

Although
 $\Ical_{\rm spec}^{\Pi}\succeq\Ical_{\rm sc}^{\Pi}$, the two information structures 
have exactly the same best sound certificate:
\begin{equation}
c_{\Ical_{\rm spec}^{\Pi}}^{\star}
=
 c_{\Ical_{\rm sc}^{\Pi}}^{\star} 
=
\lambda_{\min}(\Pi_{\rm sc}-\Pi_g).
\label{eq:app3-spectrum-equals-scalar}
\end{equation}

To see this, every matrix report $\Pi'$ compatible with either report satisfies
$\Pi'\succeq \Pi_{\rm sc}$, and therefore
\begin{equation}
\lambda_{\min}(\Pi'-\Pi_g)
\ge
\lambda_{\min}(\Pi_{\rm sc}-\Pi_g).
\end{equation}
For the full spectral report, let $v$ be a unit eigenvector associated with
$\lambda_{\min}(\Pi_{\rm sc}-\Pi_g)$ and partition it according to the node
ports as $v=[v_1^T,\ldots,v_n^T]^T$, with $v_i\in\R^2$.
The unreported eigenvector $s_i$ in \eqref{eq:app3-Q-decomposition} can always
be chosen orthogonal to $v_i$. Then
\begin{equation}
v^T(\Pi'-\Pi_{\rm sc})v
=
\sum_{i=1}^{n}r_i(s_i^Tv_i)^2=0.
\end{equation}
Thus
$v^T(\Pi'-\Pi_g)v=\lambda_{\min}(\Pi_{\rm sc}-\Pi_g)$.
By the Rayleigh--Ritz characterization,
$\lambda_{\min}(\Pi'-\Pi_g)\le v^T(\Pi'-\Pi_g)v$; together with the preceding
lower bound, equality follows. Hence \eqref{eq:app3-spectrum-equals-scalar}
holds even when every $r_i$ is known exactly. Equivalently,
\begin{equation}
 \Delta c^*(\Ical_{\rm spec}^{\Pi}\mid\Ical_{\rm sc}^{\Pi};\Sigma)=0. 
\label{eq:app3-zero-spectrum-value}
\end{equation}
This result makes the distinction between the \emph{amount} and the
\emph{stability value} of information explicit. Reporting the full spectrum
may add many exact numbers, yet it does not reduce the worst-case
conservatism unless the associated eigenvectors are also resolved.

\subsection{Stability Value of Matrix-Valued Node Information}

The full matrix report removes the rotational ambiguity retained by the scalar
and spectral reports. Therefore,
\begin{equation}
 c_{\Ical_{\rm mat}^{\Pi}}^{\star} 
=
\lambda_{\min}(\Pi-\Pi_g)
=
m(\Sigma),
\label{eq:app3-matrix-certificate}
\end{equation}
and the exact stability value of upgrading from scalar to matrix passivity
information is
\begin{equation}
 \Delta c^*(\Ical_{\rm mat}^{\Pi}\mid\Ical_{\rm sc}^{\Pi};\Sigma) 
=
\lambda_{\min}(\Pi-\Pi_g)
-
\lambda_{\min}(\Pi_{\rm sc}-\Pi_g)
\ge0.
\label{eq:app3-exact-matrix-value}
\end{equation}
This is exactly the information-induced conservatism introduced by scalarizing the matrix passivity reports into scalar ones.

Let $v_{\rm sc}$ and $v_{\rm mat}$ be unit eigenvectors associated with the
smallest eigenvalues of $\Pi_{\rm sc}-\Pi_g$ and $\Pi-\Pi_g$, respectively,
and denote their $i$-th node blocks by $v_i^{\rm sc}$ and $v_i^{\rm mat}$.
The same variational eigenvalue reasoning gives
\begin{equation}
\sum_{i=1}^{n}r_i(s_i^Tv_i^{\rm mat})^2
\le
 \Delta c^*(\Ical_{\rm mat}^{\Pi}\mid\Ical_{\rm sc}^{\Pi};\Sigma) 
\le
\sum_{i=1}^{n}r_i(s_i^Tv_i^{\rm sc})^2.
\label{eq:app3-matrix-value-bounds}
\end{equation}
Thus, the value of matrix passivity information is determined not by the
additional eigenvalue surplus $r_i$ alone, but by its alignment with the
critical stability eigenvector.

\subsection{Exact Frontier for Partial Directional Node Information}
\label{subsec:partial-direction-frontier}

The scalar--spectral--matrix comparison above gives only the two endpoints of a richer design problem.
Instead of pure scalar or pure matrix representation for all nodes, suppose only a subset of node eigendirections can be reported. Which node directions are most valuable, and by how much is the conservatism reduced? This produces an exact instance of the general frontier
in Section~\ref{subsec:general-frontier}.

Using \eqref{eq:app3-Q-decomposition}, define the embedded directional surplus
\begin{equation}
Q_i
:=
\operatorname{blkdiag}
(0,\ldots,0,\,r_i s_i s_i^T,\,0,\ldots,0)
\in\mathbb S^{2n},
\label{eq:embedded-directional-surplus}
\end{equation}
where the only nonzero $2\times2$ block is the $i$-th one. Then
\begin{equation}
\Pi
=
\Pi_{\rm sc}+\sum_{i=1}^{n}Q_i.
\label{eq:Pi-directional-sum}
\end{equation}
 For $A\subseteq[n]$, let $\Ical_A^{\Pi}$ report $\Pi_g$, all eigenvalue pairs 
$\{\pi_i,\pi_i+r_i\}$, and in addition the eigendirections $s_i$ for
 $i\in A$. Thus $A=\varnothing$ is the spectral reporting interface 
 $\Ical_{\rm spec}^{\Pi}$, whereas $A=[n]$ is the full matrix reporting interface. 
Define
\begin{equation}
 M_A 
:=
 \Pi_{\rm sc}-\Pi_g+\sum_{i\in A}Q_i. 
 \label{eq:partial-direction-MA} 
\end{equation}

\begin{proposition}[Optimal Certificate with Partially Revealed Eigendirections]
\label{thm:partial-direction-exact}
 For every $A\subseteq[n]$, the best sound certificate permitted by 
 $\Ical_A^{\Pi}$ is 
\begin{equation}
 c_{\Ical_A^{\Pi}}^{\star} 
=
 \lambda_{\min}(M_A).
\label{eq:partial-direction-exact-certificate}
\end{equation}
Consequently, at the true system $\Sigma$,
\begin{equation}
 \Delta_{\rm info}(\Sigma;\Ical_A^{\Pi}) 
=
 \lambda_{\min}(\Pi-\Pi_g)-\lambda_{\min}(M_A). 
\label{eq:partial-direction-exact-gap}
\end{equation}
\end{proposition}

\begin{proof}
 Every matrix report compatible with $\Ical_A^{\Pi}$ has the form 
\begin{equation}
 M_A+ 
 \sum_{i\notin A}\widetilde Q_i, 
\qquad
\widetilde Q_i\succeq0,
\label{eq:partial-direction-compatible}
\end{equation}
where $\widetilde Q_i$ has the same rank-one eigenvalue surplus $r_i$ in its
$i$-th block but an unresolved unit direction $\widetilde s_i$. Hence
\begin{equation}
\lambda_{\min}
\left(
 M_A+\sum_{i\notin A}\widetilde Q_i 
\right)
\ge
 \lambda_{\min}(M_A). 
\end{equation}
 Let $v$ be a unit eigenvector of $M_A$ associated with 
 $\lambda_{\min}(M_A)$ and partition it as 
 $v=[v_1^T,\ldots,v_n^T]^T$. For every unresolved node $i\notin A$, choose an 
admissible $\widetilde s_i\perp v_i$; if $v_i=0$, any unit direction may be
used. Then
 $v^T\widetilde Q_i v=r_i(\widetilde s_i^Tv_i)^2=0$ for all $i\notin A$. 
Therefore the Rayleigh quotient of the compatible completion in
\eqref{eq:partial-direction-compatible} along $v$ equals
 $\lambda_{\min}(M_A)$, which proves that the lower bound is attainable. The 
gap formula follows from \eqref{eq:app3-passivity-margin}.
\end{proof}

With unit cost per revealed node direction, define the $k$-direction
pointwise frontier by
\begin{equation}
\Phi_{\Pi}(k;\Sigma)
:=
 \min_{A\subseteq[n]:\,|A|\le k} 
 \Delta_{\rm info}(\Sigma;\Ical_A^{\Pi}). 
\label{eq:passivity-k-frontier-def}
\end{equation}
Proposition~\ref{thm:partial-direction-exact} yields the exact expression
\begin{equation}
\begin{aligned}
\Phi_{\Pi}(k;\Sigma)
=
m(\Sigma)
 -\max_{A\subseteq[n]:\,|A|\le k} 
\lambda_{\min}\!\left(
 \Pi_{\rm sc}-\Pi_g+\sum_{i\in A}Q_i 
\right).
\end{aligned}
\label{eq:passivity-k-frontier}
\end{equation}
It satisfies
\begin{equation}
\Phi_{\Pi}(0;\Sigma)
=
\Delta_{\rm info}(\Sigma;\Ical_{\rm spec}^{\Pi})
=
 \Delta_{\rm info}(\Sigma;\Ical_{\rm sc}^{\Pi}), 
\qquad
\Phi_{\Pi}(n;\Sigma)=0,
\label{eq:passivity-frontier-endpoints}
\end{equation}
and is nonincreasing in $k$. Thus the previous result that spectral
 information has zero value relative to the scalar reporting interface becomes the 
left endpoint of a complete directional-information frontier rather than an
isolated comparison.

The best certified margin at budget $k$ can be computed as the mixed-integer
semidefinite program
\begin{equation}
\begin{aligned}
\max_{x,t}\quad &t\\
\mathrm{s.t.}\quad
&\Pi_{\rm sc}-\Pi_g+\sum_{i=1}^{n}x_iQ_i\succeq tI_{2n},\\
&\sum_{i=1}^{n}x_i\le k,\qquad x_i\in\{0,1\}.
\end{aligned}
\label{eq:passivity-frontier-misdp}
\end{equation}
 Accordingly, when the full matrix reporting interface certifies a positive margin, the 
minimum number of node directions required to retain a positive certificate is
\begin{equation}
k_{\rm cert}^{\star}
:=
 \min_{A\subseteq[n]} 
 \left\{|A|: 
\lambda_{\min}
\left(
 \Pi_{\rm sc}-\Pi_g+\sum_{i\in A}Q_i 
\right)>0
\right\}.
\label{eq:minimum-directions-positive-certificate}
\end{equation}
 This quantity gives a direct interpretation for reporting-interface design: the transition 
from scalar to full matrix passivity need not be all-or-nothing; only the
node directions that move the certificate across the required margin need to
be exposed.

\subsection{Node--Network Direction Matching as Joint Information}
\label{subsec:joint-direction-information}

 The network- and node-side sources of conservatism above are two views of the same missing 
information. On the node side, the scalar passivity report replaces
$\Pi_i$ by
\begin{equation}
\Pi_{i,\rm sc}=\pi_i I_2,
\end{equation}
and discards the directional surplus
\begin{equation}
\Pi_i-\Pi_{i,\rm sc}=r_i s_i s_i^T .
\end{equation}
The conservatism introduced by scalar passivity can therefore be expressed
through the eigenvalue perturbation between the scalar and matrix certificates.
Let $v_{\rm sc}$ be the unit eigenvector associated with the smallest
eigenvalue of $\Pi_{\rm sc}-\Pi_g$, and denote its $i$-th node block by
$v_i^{\rm sc}$. By first-order eigenvalue perturbation,
\begin{equation}
\begin{aligned}
 \Delta c^*(\Ical_{\rm mat}^{\Pi}\mid\Ical_{\rm sc}^{\Pi};\Sigma)= 
\lambda_{\min}(\Pi-\Pi_g)
-
\lambda_{\min}(\Pi_{\rm sc}-\Pi_g)=
\sum_{i=1}^{n}r_i(s_i^Tv_i^{\rm sc})^2
+
O(\|\Pi-\Pi_{\rm sc}\|^2).
\end{aligned}
\label{eq:joint-matrix-information-value}
\end{equation}
Therefore, the conservatism caused by scalarization is not determined by the
discarded eigenvalue surplus $r_i$ alone. It depends on how much of this
surplus is aligned with the critical eigenvector of the scalar certificate.

For each node, the relevant directional information is
\begin{equation}
(s_i^Tv_i)^2
=
\|v_i\|^2\cos^2\theta_i ,
\label{eq:joint-direction-overlap}
\end{equation}
where $\theta_i$ denotes the relative angle between the passivity eigenvector
$s_i$ and the critical network/node-interconnection eigenvector $v_i$.
 If $v_i$ is known but $s_i$ remains unresolved in the information fiber, an 
admissible $s_i\perp v_i$ makes \eqref{eq:joint-direction-overlap} zero.
Conversely, if $s_i$ is known but the critical direction remains unresolved,
an admissible $v_i\perp s_i$ produces the same worst case. Hence one-sided
directional information has no stability value unless the relative alignment
between the two sides is resolved.

This gives a common interpretation of the two preceding applications. In Section~\ref{sec:network-spectrum-insufficiency}, $\spec(L)$ retains network strengths but removes the network eigenvectors relative to heterogeneous nodes stored in $K^{1/2}LK^{1/2}$. On the node side, a scalar or eigenvalue-only passivity report retains local strengths but removes the passivity eigenvectors relative to the critical interconnection direction. The task-relevant object in both cases is therefore not an eigenvalue or eigenvector in isolation, but their \emph{relative alignment}.

 \section{Cross-Node Information-Induced Conservatism under Block-Diagonal Compression} 
\label{sec:cross-node-info-value}

The preceding section studies the conservatism caused by compressing each
node's coupling information into a scalar report, i.e., by neglecting
\emph{intra-node} directional information. We now move outward and ask how
much additional conservatism is introduced by neglecting \emph{cross-node}
information. In particular, even if every local matrix $\Pi_i$ is known
exactly, the standard distributed passivity structure still restricts the
node-side matrix to
$\Pi=\operatorname{blkdiag}(\Pi_1,\ldots,\Pi_n)$. This section quantifies the
stability value of the cross-node information discarded by this
 block-diagonal reporting interface. 

An important distinction is needed first. Let $\Ical$ denote the information
structure under which the node side provides strictly local information.
Under independent local certification, the local facts supporting $\Pi_i$
imply the aggregate matrix $\Pi$ and nothing stronger across nodes. Thus the
block-diagonal structure is not an additional approximation once $\Ical$ has
already been fixed: it is the mathematical consequence of the local
 reporting interface itself. In this sense, the classical distributed 
passivity composition is already optimal under the predetermined $\Ical$. The question below is
instead the value of refining $\Ical$ by additional certified cross-node
quadratic information $\Jcal$, i.e., the price of decentralization measured
by Definition~\ref{def:value-information}.

Under the local information structure, the best sound passivity certificate is
\begin{equation}
c_{\Ical}^{\star}(\Sigma)
=
\lambda_{\min}(\Pi-\Pi_g).
\label{eq:cross-local-certificate}
\end{equation}

To form an information-limited benchmark, keep all local diagonal blocks $\Pi_i$ fixed. Consider a refined interface allowing full symmetric joint matrices $\widetilde{\Pi}\in \mathbb S$ with $\widetilde{\Pi}_{ii}=\Pi_i$. The best margin available under this idealized joint refinement is then 

\begin{equation}
\begin{aligned}
c_{\{\Ical,\Jcal\}}^{\star}(\Sigma)=
\sup_{\substack{
\widetilde{\Pi}\in\mathbb{S},\,\widetilde{\Pi}_{ii}=\Pi_i}}
\lambda_{\min}(\widetilde{\Pi}-\Pi_g)=
\min_i
\lambda_{\min}\!\left(
\Pi_i-(\Pi_g)_{ii}
\right).
\end{aligned}
\label{eq:cross-joint-optimum}
\end{equation}
Indeed, the $i$-th principal block of
$\widetilde{\Pi}-\Pi_g$ is always
$\Pi_i-(\Pi_g)_{ii}$, so its minimum eigenvalue cannot exceed the right-hand
side of \eqref{eq:cross-joint-optimum}. Equality is attained by choosing
$\widetilde{\Pi}_{ij}=(\Pi_g)_{ij}$ for $i\neq j$, which cancels all
cross-node blocks of the network term and leaves
$\widetilde{\Pi}-\Pi_g$ block diagonal.

Therefore, the exact stability value of the cross-node information is
\begin{equation}
\begin{aligned}
 \Delta c^*(\Jcal\mid\Ical;\Sigma) 
=
c_{\{\Ical,\Jcal\}}^{\star}(\Sigma)
-
c_{\Ical}^{\star}(\Sigma)=
\min_i\lambda_{\min}\!\left(\Pi_i-(\Pi_g)_{ii}\right)-\lambda_{\min}(\Pi-\Pi_g)\ge0.
\end{aligned}
\label{eq:cross-block-gap}
\end{equation}
\emph{Hence, the price of decentralization is not a separate algebraic quantity: in
the terminology of this paper, it is precisely the stability value of the
 cross-node relational information excluded by the local reporting interface.} Equation~\eqref{eq:cross-block-gap} is the maximum certificate 
margin that can be recovered under the joint-information refinement above.

{For compactness, define the block-diagonal and off-block parts of the network index by
$\Pi_{g,\rm bd}:=\operatorname{blkdiag}((\Pi_g)_{11},\ldots,(\Pi_g)_{nn})$
and $\Pi_{g,\rm off}:=\Pi_g-\Pi_{g,\rm bd}$.}
The role of network cross-node coupling is explicit. By the Weyl inequality,
\begin{equation}
c_{\Ical}^{\star}(\Sigma)
\ge
c_{\{\Ical,\Jcal\}}^{\star}(\Sigma)
-
\|\Pi_{g,\rm off}\|_2 .
\label{eq:cross-block-margin-lower-bound}
\end{equation}
Equivalently,
\begin{equation}
\boxed{
0\le
 \Delta c^*(\Jcal\mid\Ical;\Sigma) 
\le
\|\Pi_{g,\rm off}\|_2}.
\label{eq:cross-block-gap-bound}
\end{equation}
Thus the off-block part of the network passivity matrix gives a global upper
bound on the stability value of cross-node information. If the network has
no cross-node coupling in these quadratic coordinates, then
 $\Delta c^*(\Jcal\mid\Ical;\Sigma)=0$ and the distributed block-diagonal 
restriction is lossless.

 More generally, cross-node coupling can be nonzero while its stability value 
 is still zero. From \eqref{eq:cross-block-gap}, 
\begin{equation}
 \Delta c^*(\Jcal\mid\Ical;\Sigma)=0 
\end{equation}
if and only if
\begin{equation}
\Pi-\Pi_g
\succeq
\left[
\min_i
\lambda_{\min}\!\left(
\Pi_i-(\Pi_g)_{ii}
\right)
\right]I_{2n}.
\label{eq:cross-zero-gap}
\end{equation}
This condition has a direct directional interpretation. Suppose node $i$
attains the minimum in \eqref{eq:cross-joint-optimum}, and let $x_i$ be a
corresponding unit eigenvector of
$\Pi_i-(\Pi_g)_{ii}$. Let $x$ denote the full vector whose only nonzero block
is $x_i$ at node $i$. Whenever
 $\Delta c^*(\Jcal\mid\Ical;\Sigma)=0$, \eqref{eq:cross-zero-gap} implies 
\begin{equation}
\Pi_{g,\rm off}x=0.
\label{eq:cross-zero-gap-direction}
\end{equation}
Hence nonzero cross-node coupling does not by itself imply a positive
 stability value. If the cross-node network coupling does not act on the 
locally weakest directions and does not create a more adverse direction
elsewhere, the local information structure already attains the joint optimum,
 so that $\Delta c^*(\Jcal\mid\Ical;\Sigma)=0$. The conservatism of 
decentralization is therefore governed not only by the magnitude of the
neglected cross-node coupling, but by whether that relational information acts
along the critical local passivity directions.

\section{Computable Local Approximation of the Information Frontier}
\label{sec:quantification}

\subsection{Local Unresolved Stability Information}

 The preceding sections define the exact information-limited benchmark through an optimization over the full fiber $\Fcal_{\Ical}(z)$. Computing this optimum over the full information fiber may be expensive for large systems. When the unresolved part of the fiber is locally parameterized, the information-induced conservatism can instead be approximated from the sensitivity of the stability task. 

 Fix an information structure $\Ical$ and consider the continuous ambiguity that remains inside one of its information fibers. Let $\vartheta\in\R^n$ collect the continuous quantities not exactly resolved by the current information structure. In the applications above, such coordinates may parameterize unresolved network--node alignment, rotations of local passivity eigendirections, or missing cross-node quadratic terms. The radius $\rho_{\Ical}$ below measures the local resolution of these hidden coordinates. Locally, write the fiber as

\begin{equation}
\vartheta+\mathcal U_{\Ical},
\qquad
0\in\mathcal U_{\Ical},
\label{eq:local-fiber}
\end{equation}
and define its radius
\begin{equation}
\rho_{\Ical}
:=
\sup_{\delta\in\mathcal U_{\Ical}}\|\delta\|_2.
\label{eq:fiber-radius}
\end{equation}

\begin{definition}[Unresolved Stability Information]
\label{def:S-I}
If $m(\vartheta)$ is differentiable at the nominal point, define
\begin{equation}
\mathcal S_{\Ical}(\vartheta;m)
=
\sup_{\delta\in\mathcal U_{\Ical}} -\delta^T\nabla m(\vartheta).
\label{eq:S-I}
\end{equation}
\end{definition}
When the task and nominal point are clear, it is abbreviated as
$\mathcal S_{\Ical}$.
The quantity $\mathcal S_{\Ical}$ has the same unit as $m$. It is the largest
first-order stability-margin drop hidden by variations that the current
information cannot distinguish.

\begin{example}\label{ex:box-local-fiber}
A box-like fiber $|\delta_i|\le r_i$ yields
\begin{equation}
\boxed{
\mathcal S_{\Ical}=\sum_{i=1}^{n}
r_i\left|\frac{\partial m}
{\partial\vartheta_i}\right|.
}
\end{equation}
Thus raw uncertainty widths become comparable only after they are weighted by
their relevance to the stability boundary.
\end{example}

\begin{theorem}[Local Information--Conservatism Law]
\label{thm:local-information-law}
Assume that $m$ is twice continuously differentiable on
$\vartheta+\mathcal U_{\Ical}$ and that
\begin{equation}
\|\nabla^2m(\vartheta+\delta)\|_2\le L_m,
\qquad
\forall\delta\in\mathcal U_{\Ical}.
\label{eq:hessian-bound}
\end{equation}
Then
\begin{equation}
\mathcal S_{\Ical}
-\frac{L_m}{2}\rho_{\Ical}^2
\le
\Delta_{\rm info}(\vartheta;\Ical)
\le
\mathcal S_{\Ical}
+\frac{L_m}{2}\rho_{\Ical}^2.
\label{eq:local-information-law}
\end{equation}
In particular, as the information resolution increases such that
$\rho_{\Ical}\rightarrow0$,
\begin{equation}
\boxed{
\Delta_{\rm info}
=
\mathcal S_{\Ical}
+
O(\rho_{\Ical}^2)}.
\label{eq:first-order-gap}
\end{equation}
\end{theorem}

\begin{proof}
Let $g=\nabla m(\vartheta)$. Taylor's theorem and
\eqref{eq:hessian-bound} give, for every
$\delta\in\mathcal U_{\Ical}$,
\begin{equation}
\begin{aligned}
m(\vartheta)+g^T\delta
-\frac{L_m}{2}\|\delta\|_2^2\le m(\vartheta+\delta)\le
m(\vartheta)+g^T\delta
+\frac{L_m}{2}\|\delta\|_2^2.
\end{aligned}
\end{equation}
Taking the infimum over $\delta$ and using
\begin{equation}
\inf_{\delta\in\mathcal U_{\Ical}}g^T\delta
=
-\mathcal S_{\Ical}
\end{equation}
yields \eqref{eq:local-information-law} after substitution into
\eqref{eq:exact-gap}.
\end{proof}

Equivalently, the best certificate itself satisfies
\begin{equation}
m(\vartheta)-\mathcal S_{\Ical}
-\frac{L_m}{2}\rho_{\Ical}^2
\le
c_{\Ical}^{\star}
\le
m(\vartheta)-\mathcal S_{\Ical}
+\frac{L_m}{2}\rho_{\Ical}^2.
\label{eq:certificate-local-bounds}
\end{equation}
Define the two endpoints as
\begin{equation}
\begin{aligned}
\underline c_{\Ical}:=m(\vartheta)-\mathcal S_{\Ical}
-\frac{L_m}{2}\rho_{\Ical}^2,\;\overline c_{\Ical}:=m(\vartheta)-\mathcal S_{\Ical}
+\frac{L_m}{2}\rho_{\Ical}^2.
\label{eq:certificate-endpoints}
\end{aligned}
\end{equation}
They provide an operational three-way decision. If
$\underline c_{\Ical}>0$, the current information is already sufficient to
establish a positive sound margin. \emph{If $\overline c_{\Ical}\le0$, even the best
sound certificate $c_{\Ical}^{\star}$ cannot be positive, so no alternative
algorithm using the same information can certify stability.} If the interval
$[\underline c_{\Ical},\overline c_{\Ical}]$ crosses zero, the local bound is
inconclusive and one should refine the information or tighten the local bound.

This relation turns the abstract benchmark into an operational local test. 
It estimates how much margin is hidden by the unresolved information. It supports the same diagnosis as above: is the present reporting interface already adequate, or would resolving a stability-sensitive direction materially improve certification?

\subsection{Local Budgeted Information Allocation}
\label{subsec:local-budgeted-allocation}

The local law also gives an explicit approximation of the budgeted frontier.
Consider the box-like ambiguity in Example~\ref{ex:box-local-fiber} with
$|\delta_i|\le r_i$, and suppose that one admissible information primitive
resolves one hidden coordinate exactly. For a selected coordinate set
 $A\subseteq[n]$, the refined local fiber is 
\begin{equation}
 \mathcal U_A 
=
\left\{\delta:
 \delta_i=0\ \text{for }i\in A,\quad 
 |\delta_i|\le r_i\ \text{for }i\notin A 
\right\}.
\label{eq:local-refined-box}
\end{equation}
Define the first-order value score
\begin{equation}
a_i
:=
r_i\left|\frac{\partial m}{\partial\vartheta_i}\right|,
\qquad
a_{(1)}\ge a_{(2)}\ge\cdots\ge a_{(n)},
\label{eq:local-information-score}
\end{equation}
where $(\cdot)$ denotes the decreasing ordering.

\begin{proposition}[First-Order $k$-Information Frontier]
\label{cor:local-k-frontier}
Let $\rho_0^2:=\sum_{i=1}^{n}r_i^2$ and assume the Hessian bound in
Theorem~\ref{thm:local-information-law} holds uniformly on the baseline box.
Under a unit-cost budget that resolves at most $k$ coordinates, define
\begin{equation}
\Phi_{\rm loc}(k;\vartheta)
:=
 \min_{A\subseteq[n]:\,|A|\le k} 
 \Delta_{\rm info}(\vartheta;\Ical_A). 
\label{eq:local-k-frontier-def}
\end{equation}
Then
\begin{equation}
\boxed{
\Phi_{\rm loc}(k;\vartheta)
=
\sum_{j=k+1}^{n}a_{(j)}
+O(\rho_0^2).}
\label{eq:local-k-frontier}
\end{equation}
Hence, to first order, the optimal information policy is to resolve the $k$
coordinates with the largest scores
$r_i|\partial m/\partial\vartheta_i|$.
\end{proposition}

\begin{proof}
For the refined fiber \eqref{eq:local-refined-box}, Definition~\ref{def:S-I}
gives
\begin{equation}
 \mathcal S_{\Ical_A} 
=
 \sum_{i\notin A}a_i. 
\label{eq:local-S-selected}
\end{equation}
 The smallest first-order term over all $|A|\le k$ is therefore obtained by 
removing the $k$ largest $a_i$, yielding
$\sum_{j=k+1}^{n}a_{(j)}$. Moreover every refined radius satisfies
 $\rho_{\Ical_A}\le\rho_0$. Applying 
 Theorem~\ref{thm:local-information-law} uniformly to all admissible $A$ shows 
that optimizing the exact gap can change this minimum only by a term bounded
in magnitude by $L_m\rho_0^2/2$, which proves
\eqref{eq:local-k-frontier}.
\end{proof}

Equation~\eqref{eq:local-k-frontier} gives a simple operational rule that is
absent from an unweighted information count: a coordinate is valuable when it
is both poorly resolved and strongly aligned with the stability gradient. With
unequal primitive costs $w_i$, the first-order design becomes the budgeted
selection problem
\begin{equation}
 \max_{A\subseteq[n]} 
 \sum_{i\in A}a_i 
\qquad
\mathrm{s.t.}\qquad
 \sum_{i\in A}w_i\le B, 
\label{eq:local-weighted-selection}
\end{equation}
which is the local counterpart of the general frontier
\eqref{eq:information-frontier}. Section~\ref{subsec:general-frontier}, Section~\ref{subsec:partial-direction-frontier}, and the present section address the same design-oriented question: \emph{which additional information gives the largest reduction in irreducible conservatism under a limited reporting-interface budget?}

\section{Conclusion}
\label{sec:conclusion}
 This paper develops a method-independent framework for understanding conservatism in distributed stability certification from the perspective of information. For a prescribed information structure, the best sound certificate is determined by the worst stability margin among all systems that the reporting interface cannot distinguish. This benchmark separates the total conservatism of a particular certificate into information-induced conservatism and method-induced conservatism, and converts information sufficiency into a property of the information fibers induced by the information structure. In particular, indistinguishable systems with materially different stability margins provide constructive witnesses of a limitation that no alternative certificate using the same information can overcome. 

The network--node studies reveal a common mechanism behind several apparently different forms of conservatism. 
Across the three studies, compression removes stability-relevant relations, including network--node alignment, local eigendirections, and cross-node coupling information.
Adding quantities that do not recover these relations need not reduce certification conservatism.
Therefore, the stability value of information is governed not by the number of reported quantities, but by whether the information preserves the relations along which the stability task varies. 

The information--conservatism frontier extends this diagnosis into a reporting-interface-design problem by quantifying the best achievable conservatism under restricted admissible information. Its inverse characterizes the minimum information required to attain a prescribed accuracy, while the witness-pair formulation identifies exactly which task-relevant ambiguities must be resolved. The exact directional frontier for matrix passivity reports and the local sensitivity-based information allocation further demonstrate how this principle can be made constructive. The resulting framework provides a systematic basis for deciding where additional complexity is worthwhile to reduce conservatism. If the available information is underused, the analysis method should be improved; if the information-limited benchmark itself is restrictive, the reporting interface should be enriched.

\section*{Acknowledgment}
The authors used OpenAI ChatGPT to assist with the coding of numerical validation and with language editing throughout the manuscript. All AI-assisted code and text were reviewed and verified by the authors.